\documentclass[%
superscriptaddress,
 amsmath,amssymb,
 aps,
longbibliography
]{revtex4-2}

\usepackage{graphicx}
\usepackage{dcolumn}
\usepackage{bm}

\usepackage{xcolor}

\usepackage[normalem]{ulem}

\begin{document}

\preprint{APS/123-QED}

\title{CO$_2$ convective dissolution in a 3-D granular porous medium:\\ an experimental study}

\author{Christophe Brouzet}
\email{christophe.brouzet@univ-amu.fr}
\altaffiliation[Present address: ]{Universit\'e C\^ote d'Azur, CNRS, Institut de Physique de Nice, 06100 Nice, France}
\affiliation{Aix Marseille Univ, CNRS, Centrale Marseille, IRPHE - Marseille, France}%

\author{Yves M\'eheust}
\affiliation{Univ. Rennes, CNRS, G\'eosciences Rennes - UMR 6118, 35000 Rennes, France}%


\author{Patrice Meunier}
\affiliation{Aix Marseille Univ, CNRS, Centrale Marseille, IRPHE - Marseille, France}%

\date{\today}

\begin{abstract}
Geological storage of CO$_2$ in deep saline aquifers is a promising measure to mitigate global warming by reducing the concentration of this greenhouse gas in the atmosphere. When CO$_2$ is injected in the geological formation, it dissolves partially in the interstitial brine, thus rendering it denser than the CO$_2$-devoid brine below, which creates a convective instability. The resulting convection accelerates the rate of CO$_2$ perennial trapping by dissolution in the brine. The instability and resulting convection have been intensively discussed by numerical and theoretical approaches at the Darcy scale, but few experimental studies have characterized them quantitatively. By using both refractive index matching and planar laser induced fluorescence, we measure for the first time the onset characteristics of the convective dissolution instability in a 3-D porous medium located below a gas compartment. Our results highlight that the dimensional growth rate of the instability remains constant when the CO$_2$ partial pressure in the compartment is varied, in clear discrepancy with the theoretical predictions. Furthermore, within the CO$_2$ partial pressure range studied, the measured growth rate is one to three orders of magnitude larger than the predicted value. The Fourier spectrum of the front is very broad, highlighting the multi-scale nature of the flow. Depending on the measurement method and CO$_2$ partial pressure, the mean wavelength is $1$ to $3$ times smaller than the predicted value. Using a theoretical model developed recently by Tilton (J.~Fluid Mech., vol.~\textbf{838}, 2018), we demonstrate that these experimental results are consistent with a forcing of convection by porosity fluctuations. Finally, we discuss the possible effects of this forcing by the porous medium's pore structure on the CO$_2$ flux across the interface, measured in these experiments about one order of magnitude higher than expected. These results obtained in model laboratory experiments show that accounting for sub-Darcy scale  flow heterogeneities may be necessary to correctly predict convective dissolution during CO$_2$ subsurface sequestration.
\end{abstract}

\maketitle


\section{Introduction\label{sec:level1}}

Over the past ten years, more than $30$~gigatonnes of carbon dioxide (CO$_2$) have been released annually into the atmosphere due to human activities~\cite{IEA2021}. Due to this very large release rate, and as CO$_2$ is a gas with significant capacity for greenhouse effect and significant residence time in the atmosphere, its contribution to the global warming of the Earth's atmosphere  amounts to two thirds of the global greenhouse effect~\cite{Bryant1997}. Its accumulation in the atmosphere thus results in a potentially disastrous global problem. One solution to mitigate this issue is to reduce the anthropogenic emissions to the atmosphere by capturing and storing CO$_2$ in deep saline aquifers and depleted oil reservoirs~\cite{Metzetal2005}. These geological formations, located between~$1$ and~$3$~km beneath the Earth's surface, are typically porous and filled with brine. At these depths, the CO$_2$ becomes supercritical and, once injected, its motion through the reservoir is controlled by fluid mechanics~\cite{HuppertNeufeld2014}. Being positively buoyant, it first rises to the top of the reservoir,  until it encounters an impermeable cap-rock, along which it then spreads horizontally. This leads to the formation of a supercritical CO$_2$ layer positioned above the brine. By dissolving into the surrounding brine, CO$_2$ then densifies it locally and creates a negatively buoyant CO$_2$-enriched brine layer sitting on top of pure brine. Once this new layer is sufficiently thick, it becomes gravitationally unstable, leading to a convective instability with a typical fingering pattern. This convective dissolution process is of paramount importance for CO$_2$ storage, as (i) the dissolution of CO$_2$ into brine reduces the long-term risks of CO$_2$ leakage to more superficial geological formations, and (ii) the convection continuously brings CO$_2$-devoid brine in contact with the supercritical CO$_2$, thus contributing to enhancing the dissolution rate~\cite{Neufeldetal2010,Pauetal2010,HuppertNeufeld2014,Slim2014,EmamiMeybodietal2015}. 

The convective instability is classically known to be controlled by the Rayleigh number~Ra~\cite{HuppertNeufeld2014}, which is defined in detail further in the text. The convective dissolution instability has been extensively studied using a combination of theoretical and numerical approaches at the Darcy's (i.e., continuum) scale, with a description of flow based on Darcy's law~\cite{EnnisKingetal2005,Riazetal2006,Hassanzadehetal2007,Pauetal2010,EleniusJohannsen2012,Slim2014,EmamiMeybodietal2015,DePaolietal2016,DePaolietal2017}. These studies focus principally on the onset time of the instability and its main characteristics, such as growth rate or wavelength, but also on the CO$_2$ dissolution rate. They have recently established a comprehensive scenario for the convective dissolution process, with several steps appearing as a function of time~\cite{Pauetal2010,Slim2014,DePaolietal2017}. First, the process is initiated by a \textit{diffusive regime}, in which diffusion of CO$_2$ in the brine at the top of the reservoir thickens the CO$_2$-enriched brine layer located above the pure brine. Once sufficiently dense fluid is accumulated in this diffusive layer, perturbations start to grow linearly in the \textit{linear-growth regime}. This typically creates CO$_2$-rich brine fingers going downwards and increasing the CO$_2$ flux once they are large enough, in the \textit{flux-growth regime}. Then, the fingers start to interact with their neighbors, leading to the \textit{merging regime}. This is followed by a \textit{constant-flux regime}, where the merged fingers are sufficiently spaced, allowing for a new destabilization of the diffusive boundary layer between them. Finally, when the fingers feel the influence of the bottom of the reservoir, the convective dissolution process progressively stops in what is termed the \textit{shut-down regime}. Note that these regimes do not depend directly on the Rayleigh number, except the last one. Indeed, the Rayleigh number only controls how many of these regimes can occur before the influence of the bottom of the reservoir on the process starts to be significant~\cite{Slim2014}. Note also that the predictions from Darcy scale theoretical developments are not necessarily expected to fully hold when the Darcy number based on the vertical size of the porous medium is sufficiently large so that the most unstable wavelength it not much larger than the typical pore size. Pore scale heterogeneities are then expected to play a significant role~\cite{EleniusJohannsen2012,HuppertNeufeld2014,EmamiMeybodi2017,Tilton2018}. Heterogeneities at scales larger than the Darcy scale are also expected to impact the convection significantly, but have not yet been much investigated.

These theoretical and numerical works have been completed by experimental studies, which, however, remain more limited~\cite{EmamiMeybodietal2015}. Indeed, a large part of them have been performed in a Hele-Shaw cell, mimicking Darcy's law in an experimental setup without  grains or porous structure ~\cite{KneafseyPruess2010,KneafseyPruess2011,Backhausetal2011,Slimetal2013,Tsaietal2013,Seyyedietal2014,Thomasetal2015,Vremeetal2016,Thomasetal2018}. These first experiments in Hele-Shaw cell have mainly remained qualitative~\cite{KneafseyPruess2010,KneafseyPruess2011,Seyyedietal2014}, and the validation of the theoretical predictions for the instability characteristics~\cite{Vremeetal2016} or for the convective dissolution process scenario~\cite{Slimetal2013} have only been reported recently. Some experimental works have focused on CO$_2$ flux measurements in PVT (pressure, volume and temperature) cells partly filled by a porous medium at the bottom~\cite{KneafseyPruess2011,Seyyedietal2014,NazariMoghaddametal2012,NazariMoghaddametal2015}. However, the configuration of PVT cells do not allow for visualisation of the instability within the considered medium~\cite{EmamiMeybodietal2015}. A few works have visualized the instability within a porous medium confined between two plates. However, they did not focus directly on the instability characteristics, as they remained mainly qualitative~\cite{Agartanetal2015} or as they investigated the scaling relation between the CO$_2$ flux and the Rayleigh number~\cite{Tsaietal2013,Neufeldetal2010}. Note that the instability in a porous medium has also been observed using X-ray tomography~\cite{Wangetal2016}, with a main focus on the structure of the fingering pattern. Finally, it is worth mentioning the two works of MacMinn \textit{et al.}~\cite{MacMinnetal2012,MacMinnJuanes2013} that report experiments and visualisation of the instability in a porous medium when the CO$_2$ layer at the top is not stationary but still migrates at the top of the reservoir. This configuration may be relevant for geological sequestration sites but it does not represent the classical model system reported in the literature and studied here. 

In any case, the validation of the characteristics of the instability and quantitative characterization of the convective dissolution process, obtained in the theoretical and numerical approaches, remains to be performed in an experience involving a flow cell containing a porous/granular structure. If one considers a granular porous medium, the presence of the grains may render the convective dissolution process  significantly more complex by introducing several ingredients, not considered when assuming Darcy's law in an isotropic and homogeneous continuous medium. First, as a natural porous medium is intrinsically random, it may contain heterogeneities, i.e. variations in porosity and permeability. These heterogeneities may impact the velocity field and thus the concentrations, a phenomenon which may continuously force perturbations~\cite{EmamiMeybodietal2015,Tilton2018}. Furthermore, pore scale flow heterogeneities (i.e., below the Darcy scale), may play a similar, but more subtle, role in the instability development. Secondly, the heterogeneity of the solute-advecting pore scale flow (in the pores/channels between the grains) and its interaction with molecular diffusion below the Darcy scale are known to induce hydrodynamic dispersion of the solute at the Darcy scale~\cite{Souzyetal2020}. This process enhances mixing and can be expected to  lead to coarsening of the fingering pattern~\cite{Wangetal2016,Liangetal2018}. Theoretical, numerical, and experimental works have started to study the additional effects of heterogeneities~\cite{EnnisKingetal2005,Agartanetal2015,DePaolietal2016,Salibindlaetal2018}, porosity fluctuations~\cite{Tilton2018,Pauetal2010} and hydrodynamic dispersion~\cite{HidalgoCarrera2009,EmamiMeybodietal2015,Wangetal2016,EmamiMeybodi2017,EmamiMeybodi2017b,Wenetal2018,Liangetal2018,DePaoli2021} on the convective dissolution process. In particular, they have shown that heterogeneities can reduce the onset time of the instability. However, the question of whether the description based on Darcy's law is relevant, when the wavelength of the instability is on the order of the grain size and/or when pore scale flow heterogeneities and hydrodynamic dispersion are present, remains open.

In this paper, we study the onset of convective dissolution in a granular porous medium quantitatively, in order to compare the measurements with previous theoretical and numerical predictions of the instability characteristics~\cite{EnnisKingetal2005,Riazetal2006,Hassanzadehetal2007,EleniusJohannsen2012}. The paper is organized as follow. In Section~\ref{sec:methods}, we present the experimental setup that allows us to both measure the characteristics of the instability within the porous medium and obtain the CO$_2$ flux across the liquid-gas interface. We then present a brief overview of the convective dissolution process observed in the 3-D porous medium (Section~\ref{sec:overview}), before discussing in detail the experimental results obtained on the instability characteristics (growth rate and wavelength), in Section~\ref{instability_charac}. The results for the growth rate show a clear discrepancy with the theoretical predictions, but can however be explained by a forcing mechanism based on porosity fluctuations as in a model recently introduced by Tilton~\cite{Tilton2018}. Finally, in Section~\ref{sec:Flux}, we present and discuss the experimental results obtained for the CO$_2$ flux across the interface, which appears much larger than expected.

\section{Materials and methods\label{sec:methods}}

\subsection{Experimental setup}

\begin{figure}[b!]
	\begin{centering}
		\includegraphics[width=0.65\textwidth]{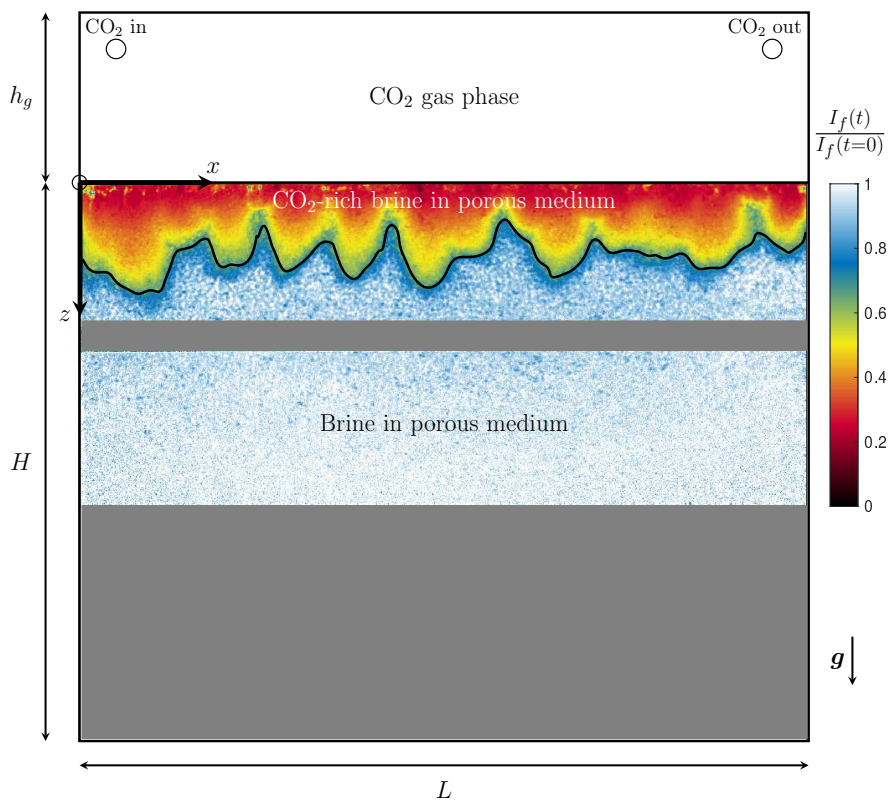}
		\caption{Sketch of the experimental setup. The porous medium, filled with brine, has a height~$H$ and the gas phase an height~$h_g$. The thickness of the tank is~$e=15$ or $7$~mm, its length is~$L=300$~mm, with a total height~$H+h_g=300$~mm. The gray parts in the image correspond to zones for which no useful data is available on the recorded images: the thin rectangle corresponds to a lateral reinforcement of the tank which prevents imaging the porous medium, while the bottom of the tank is out of the field of view of the camera, which is not an issue as we mainly focus on the onset of the instability. The experimental image inserted in the sketch and its colorbar are discussed in Section~\ref{PLIF}.\label{fig:setup}}
	\end{centering}
\end{figure} 

The experimental setup consists of a $300$~mm large, $300$~mm high, $7$ or $15$~mm thick, closed and transparent tank as shown in Fig.~\ref{fig:setup}. Note that the tank thickness~$e$ is much smaller than its dimensions in the other directions. In addition, as shown further in the text, the thickness is also much smaller than the typical wavelength of the instability but remains larger than the typical grain sizes used in the experiments. This ensures that the instability remains two-dimensional and allows us to visualise it using a refractive index matching method by looking through a relatively thin but still 3-D porous medium. As in several studies reported in the literature~\cite{KneafseyPruess2010,KneafseyPruess2011,Seyyedietal2014,Thomasetal2015,Vremeetal2016,Thomasetal2018}, we have chosen to use gaseous CO$_2$ and water as a fluid pair to model the instability in the experimental setup. Note that the physics of the process is completely similar when supercritical CO$_2$ is used (instead of gaseous CO$_2$)~\cite{Khosrokhavaretal2014}, but it is necessary to reach much higher pressures experimentally. The bottom half of the tank is therefore filled with a porous medium and salt water while the top half contains air at atmospheric pressure and a CO$_2$ sensor. The salt dissolved in the water is here employed to match the refractive index of the grains. The $z$-direction points downward and its origin $z=0$ is set at the interface between the porous medium and the air. At time~$t=0$, part of the air in the top half of the tank is removed and replaced by CO$_2$, leading to a sudden increase of the fraction $X_{CO_2}$ of CO$_2$ above the porous medium. In this configuration, the density difference~$\Delta \rho$ between CO$_2$-rich brine and pure brine is therefore proportional to the CO$_2$ partial pressure~$P_{CO_2}= X_{CO_2} P_0$ with $P_0=1$\,bar (see Section~\ref{sec:equations}). Note that during the CO$_2$ injection, the total pressure increases by less than $0.01$~bar during less than $1$~second. This adiabatic compression creates an increase of the temperature of the gas that is smaller than 3 degrees, which leads to a transient decrease of the density, smaller than~$0.1\,\%$. However, this density variation is too small to modify the convection due to CO$_2$ since its duration is several orders of magnitude shorter.

In order to visualise the instability inside the porous medium, we use the combination of two different techniques: refractive index matching to see through the porous medium and planar laser induced fluorescence to qualitatively detect CO$_2$ iso-concentrations. These techniques are presented in the next two sections.

\subsection{Porous medium and refractive index matching} \label{sec:PorousMedium}

\begin{figure}[b!]
	\begin{centering}
		\includegraphics[width=1\textwidth]{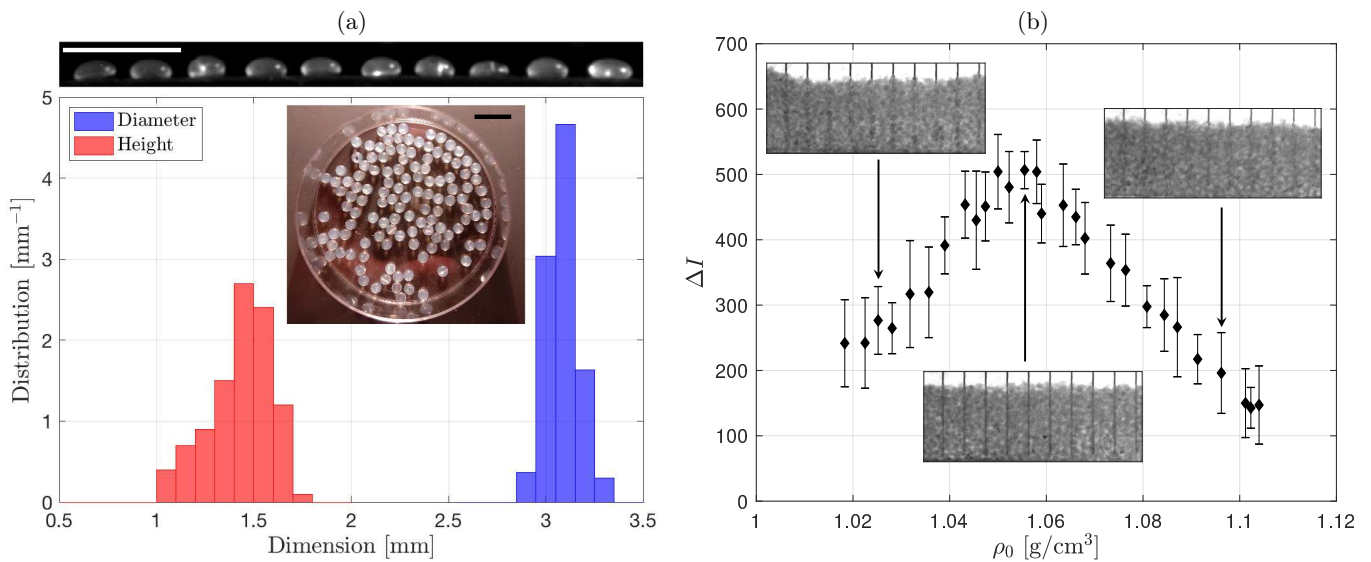}
		\caption{(a)~Histograms of grain typical dimensions. The two insets show a side-view (top) and a top-view (middle). The scale bars both indicate~$10$~mm. (b)~Contrast~$\Delta I$ as a function of the salt water density~$\rho_0$. The three insets show examples of images at different water densities obtained using a white backlight. The spacing between the lines is~$10$~mm and the error bars are computed from the standard deviation of the set of $\Delta I$~measurements obtained for the different lines in the background.\label{fig:index_matching}}
	\end{centering}
\end{figure} 

Refractive index matching allows for optical access to the bulk of dense suspensions and porous media~\cite{Dijksmanetal2012}, by matching the optical indices of the fluid and solid phases and thus canceling any light refraction at solid-liquid interfaces within the flow cell. For the refractive indices of these two phases to be identical, they have to be chosen carefully. Most of the previous works have used highly viscous fluids~\cite{Dijksmanetal2012,Souzyetal2017,DalbeJuanes2018}, with a dynamic viscosity~$\mu$ several orders of magnitude higher than that of  water. In order to reach high Rayleigh numbers in our experiments (see Section~\ref{sec:equations}) and to stay rather simple with respect to the CO$_2$ dissolution chemistry, we use FEP (fluorinated ethylene propylene) transparent particles, with a refractive index~$n=1.344$ close to water so that it can be matched with salt water. The FEP particles look like  droplets solidified while resting on a surface, as seen in Fig.~\ref{fig:index_matching}(a). Indeed, they all have a flat side (located at the bottom of the grains in the top inset in Fig.~\ref{fig:index_matching}(a)) and, at the opposite, a more rounded side. The height of the grains, defined as the distance between these two sides, is on average equal to~$1.43$~mm. Viewed from the top, they exhibit a clear circular cross section, with a mean diameter of~$3.08$~mm. The distributions of diameter and height are represented in Fig.~\ref{fig:index_matching}(a) and show that their dimensions are relatively uniform over the particle population, despite a non-classical shape whose only symmetry is cylindrical. With such grain dimensions, the tank contains therefore between $100$ and~$200$ grains in its width, $50-100$~grains in its height, and $5$ to $10$~particles in its thickness. This is sufficient to consider the assembly of grains as a three-dimensional (3-D) porous medium. Moreover, as the thickness of the tank is about $5$ to $10$~particles, the flow is fully 3-D in the bulk of the porous medium. The packing of FEP particles has a porosity~$\phi=0.39\pm0.02$ and a permeability~$K=(9.3\pm0.8)\times 10^{-10}$~m$^2$ (see Appendix~\ref{permeability} for details on the permeability measurements). Note that water does not wet FEP significantly. As a consequence, air bubbles are easily trapped in the porous medium when the grains are put together in water. To remove this issue, before the start of the experiments, we use a vacuum pump to decrease the air pressure in the gas compartment. It dilates the air bubbles trapped in the medium so that they detach more easily and rise to the surface.

Refractive index matching is performed by adding NaCl salt to the water, in order to reach the FEP's refractive index and make the porous medium as transparent as possible. Note that the presence of salt in the water has a stabilizing influence on the instability~\cite{Loodtsetal2014b,Thomasetal2018}. However, adding salt is necessary to match the refractive index, and the concentrations considered remain limited to~$1.5$~mol/l. In addition, the added salt has been chosen as pure as possible, to limit the presence of other chemical species in the water that may affect the global pH when CO$_2$ dissolves. The optimal salt concentration, or salt water density~$\rho_0$, has been found by examining the contrast on a pattern composed of vertical dark lines placed behind the tank and under a back-light exposure. The three insets in Fig.~\ref{fig:index_matching}(b) show typical images obtained in this configuration: on the top, the lines are clearly visible as there is no grains and only salt water. On the bottom, the lines may be distinguished or not, depending on the density of the salt water. The contrast~$\Delta I$, defined as the difference between the intensity in the regions without lines in background and the intensity along the lines in the background, is shown as a function of the salt water density in Fig.~\ref{fig:index_matching}(b). It clearly exhibits an optimal value for the density, in the range $\rho_0\approx 1.050-1.060$~g/cm$^3$, corresponding to the inset where the lines are the most distinguishable. This calibration has been performed using a white backlight, but we assume that the optimal salt density is the same at the wavelength of the laser (in the visible range, see Section~\ref{PLIF}) used in our experimental setup. Note that the porous medium is not completely transparent here, as the grains are not fully transparent and contribute to light attenuation. Such a combination of FEP grains and salt water may therefore not be sufficient for refractive index matching in a 3-D porous medium with large dimensions in all directions, but remains perfectly adapted to the setup described here, where the tank is much thinner in one direction. In the experiments, the salt water has been prepared with a density equal to  $\rho_0 =  1.056 \pm 0.002$~g/cm$^3$.

\subsection{Planar laser induced fluorescence\label{PLIF}}

The presence of dissolved CO$_2$ is visualised using planar laser induced fluorescence. The interstitial fluid, i.e. salt water, contains fluorescein at a uniform concentration of $C_f=10^{-5}$~mol/l. When CO$_2$ dissolves in the interstitial fluid, it decreases the pH locally, which decreases the re-emitted intensity, as fluorescein light emission and absorption are pH-dependent~\cite{MartinLindqvist1975,Walker1987,DiehlMarkuszewski1989,KlonisSawyer1996,CoppetaRogers1998} (see details in Appendix~\ref{pH}). This property of fluorescein has been used to assess pH variations in a wide range of systems, such as acid turbulent jets~\cite{Lacassagneetal2018}, CO$_2$ bubble rising in a quiescent fluid~\cite{Valiorgueetal2013}, CO$_2$~gas dissolution in turbulent water~\cite{Lacassagneetal2017}, or convective dissolution of CO$_2$ in a Hele-Shaw cell~\cite{Vremeetal2016}.

The tank is illuminated from the top, using a $500$~mW~laser with a wavelength~$\lambda_e=473$~nm (Laser Quantum, gem~$473$). From this laser, a laser sheet is obtained with a Powell lens to ensure a homogeneous intensity within the sheet. The intensity of the laser is set to $220$~mW and this quantity is kept stable during all the duration of the experiments (typically a few hours) by proper cooling of the laser. No photo-bleaching effect has been noticed during a long exposure of the fluorescein solution to such conditions. The tank is filmed from its largest side using a $16$-bit color camera (Nikon D200) equipped with a green light filter centered around the fluorescein's emission wavelength,~$\lambda_f=515$~nm. As the development of the instability is relatively slow, the frame rate is set to $1$~image per minute, while the experiments typically last several hours.

We describe here qualitatively the fluorescence intensity recorded by the camera and resulting from light emission by the fluorescein occupying the pore space of the medium. First, the intensity incoming from the laser is attenuated by the porous medium containing the solution. The intensity received at a point $M(x,z)$ of the medium at time~$t$ is given by 
\begin{equation}
I_r(x,z,t)=I_0 \exp\left(-\int_0^{z} \epsilon(x,z',t)\textrm{d}z'\right),
\end{equation}
where $I_0$ is the intensity of the laser before entering in the porous medium, and~$\epsilon$ is the absorption coefficient of the porous medium. Note that this coefficient depends on several factors. Indeed, the grains are not perfectly transparent and therefore attenuate laser intensity, the fluorescein in the pores has its own absorption coefficient (see Appendix~\ref{pH}) that has to be multiplied by the fluorescein concentration~$C_f$, and, in addition, the refractive index matching between the grains and the fluid is not perfect and may contribute to decreasing the transmitted intensity at the fluid-solid interface. Due to the pH-dependency of the fluorescein's fluorescence, this absorption coefficient depends also on pH, and therefore on time and space coordinates as the CO$_2$ concentration is inhomogeneous in the tank during an experiment. However, the variation of 
$\int_0^{z} \epsilon(x,z',t)\textrm{d}z'$ between the beginning of the experiment and the end of the experiment is small since $ \epsilon$ does not change by more than $10\,\%$ between the value measured without CO$_2$ and the one measured with the saturated concentration in CO$_2$. The difference between $I_r(x,z,t)$ and $I_r(x,z,t=0)$ is thus smaller than $10\,\%$ close to the surface (i.e. for $z\leq30$~mm). We have thus neglected this effect and considered that $I_r(x,z,t)=I_r(x,z,t=0)$.

The re-emitted intensity, collected by the camera, is therefore given by~\cite{Lacassagneetal2018}
\begin{equation}
I_f(x,z,t)= f(pH(x,z,t)) \ I_r(x,z,t=0),
\end{equation}
where the fluorescence~$f(pH)$ is a function that has been calibrated (see Appendix~\ref{pH}). The pH dependency on the CO$_2$ concentration in water, $C$, can be approximated as
\begin{equation}
pH = \frac{1}{2}  [ pK_a - \log_{10}(C) ],
\end{equation}
with $pK_a=6.37$. CO$_2$ concentrations in the medium may therefore be obtained using the ratio~\cite{Lacassagneetal2018}
\begin{equation}
\frac{I_f(x,z,t)}{I_f(x,z,t=0)}=\frac{f(pH(x,z,t))}{f(pH(x,z,t=0))}\equiv f_0(pH(x,z,t)).\label{eq:norm_int}
\end{equation}
Although the global features of the function $f(pH)$ can be fairly well explained theoretically (see Appendix~\ref{pH}), the multiplicative pre-factor~$f(pH(x,z,t=0))$ may vary weakly due to experimental conditions (fluorescein concentration, temperature, salt impurities, etc.). As a consequence, we were not able to reach a quantitative determination of the CO$_2$ concentration since a variation of only $5\,\%$ of the function~$f_0(pH)$ leads to a change in $C$ of one order of magnitude. Indeed, $C$~is extremely sensitive to the pH (since $C \sim 10^{-2pH}$) and the pH strongly depends on the fluorescence function (since $1/f_0'(pH)\approx 11$ within the range $pH=4$ to~$6$, i.e. the pH region corresponding to the saturated concentrations for the CO$_2$ partial pressures considered in this study).

However, this method can be used to detect pH iso-curves~\cite{Thomasetal2015} and therefore iso-concentration lines of CO$_2$. An example showing the ratio $I_f(x,z,t)/I_f(x,z,t=0)$ is given in Fig.~\ref{fig:setup} where the iso-concentrations are clearly visible. The iso-concentration $z_F(x,t)$ corresponding to $I_f(t)/I_f(0)=0.65$ is shown with a solid black line. Iso-$C$ fronts are extracted at each time and then smoothed using a spline function in order to remove the small scale fluctuations due to the texture of the grains. The choice of the iso-concentration was varied between~$0.65$ and~$0.85$, which slightly modifies the value of the growth rates and wavelength. These uncertainties are taken into account in the error bars of the figures presented further in the text.

\subsection{Direct measurement of CO$_2$ dissolution flux\label{flux_measure}}

To measure the fraction of CO$_2$ in air, $X_{CO_2}$, during the experiments, the mixture of air and CO$_2$ above the porous medium is extracted from the cell by a small diaphragm pump (Boxer, 22K series), brought in a CO$_2$ sensor (GSS, ExplorIR-W-F-100) and then re-injected in the cell. The positions of the hoses for this extraction/injection process are indicated in Fig.~\ref{fig:setup}. The sensor measures the fraction of CO$_2$ in the gas twice every second and the pump has a flow rate of~$1.4$~l/min, allowing a response time of the sensor  smaller than~$5$~s, according to the manufacturer. The pump-flow rate is sufficiently large for the full gas volume to be circulated through the sensor within a maximum of~$30$~s, which is small enough with respect to the typical scale of time variations of the CO$_2$ fraction in air during the experiment. This setup permits to accurately measure the CO$_2$ partial pressure throughout the experiment. 

Since the partial pressure decreases by about $10$ to $60\,\%$ during the experiments, it is possible to deduce the flux of CO$_2$ absorbed into the fluid. However, it should be noted that some CO$_2$ is lost due to the non gas-proof junctions of the tubing and the cell. This loss has been measured by replacing the water and the porous medium with a solid plate. It causes an exponential decay, $e^{-t/\tau}$, of the CO$_2$ partial pressure  with a decaying time $\tau\approx 15.6$~h. When the water and porous medium are present, the partial pressure decreases about twice faster due to absorption by the fluid.
We can thus calculate the flux of dissolved CO$_2$ using the formula 
\begin{equation}
F=-\frac{h_g}{RT}\left( \frac{\textrm{d}P_{CO_2}}{\textrm{d}t}+\frac{P_{CO_2}}{\tau}\right),\label{eq:def_flux}
\end{equation}
where $h_g$ is the height of the gas volume above the porous medium (see Fig.~\ref{fig:setup}), $R=8.314$~J/K/mol the ideal gas constant,  and $T\approx293$~K the temperature of the gas. In order to limit the effect of noise, the partial pressure~$P_{CO_2}$ is first interpolated as a function of time with a spline function before its time derivative is computed.

As the flux measurements are purely based on the decay of the CO$_2$ partial pressure in the gas compartment, it is fully possible to use other granular materials that FEP, as was done in several studies in PVT cells reported in the literature~\cite{KneafseyPruess2011,NazariMoghaddametal2012,Seyyedietal2014,NazariMoghaddametal2015}. We have therefore performed a few additional experiments with other grains, which have been chosen spherical and with a typical dimension (here the diameter~$d$) smaller than that of the FEP grains. Their characteristics are given in Table~\ref{table:extra_grains}, for comparison with the FEP grains. As these grains are made of different materials, they are not adapted at all to the refractive index matching technique with salt water, and the quantitative visualisation of the instability shown in Fig.~\ref{fig:setup} is not possible. However, these grains allow us to explore smaller permeabilities (also measured using the method described in Appendix~\ref{permeability}) compared to the FEP grains and to observe the dependence of the CO$_2$ flux on this parameter. The results obtained with these grains are only discussed in Section~\ref{sec:Flux}.

\begin{table}[h!]
\caption{Table with the characteristics of the different grains used in this study. 
\label{table:extra_grains}
}
\begin{ruledtabular}
\begin{tabular}{lllllll}
material & shape & $d$~[mm] & $\phi$ & $K$~[m$^{2}$] & number of exp. & measured quantities\\
\colrule
FEP & half-sphere & $1.5$ & $0.37-0.41$ & $9.30\pm0.80 \times 10^{-10}$ & $20$ & growth rate, wavelength, flux \\
PMMA & sphere & $1$ & $0.39$ & $2.69\pm0.02 \times 10^{-10}$ & $1$ & flux only \\
silica & sphere & $0.35$ & $0.43-0.45$ & $0.94\pm0.07 \times 10^{-10}$ & $2$ & flux only\\
\end{tabular}
\end{ruledtabular}
\end{table}

\subsection{Governing equations and non-dimensional parameters\label{sec:equations}}

Within the porous medium, the classical approach assumes that the flow is governed by three equations~\cite{EnnisKingetal2005,Riazetal2006,Hassanzadehetal2007,EleniusJohannsen2012,Slim2014,EmamiMeybodietal2015,EmamiMeybodi2017,EmamiMeybodi2017b,Tilton2018}:
\begin{eqnarray}
\textrm{fluid incompressibility:}&~~~&
\bm{\nabla} \cdot \bm{v}=0,\label{eq:incomp}\\
\textrm{Darcy's law:}&~~~&\mu \bm{v}=-K(\bm{\nabla}P-\Delta \rho \, c \, \bm{g}),\label{eq:Darcy}\\ 
\textrm{advection-diffusion equation:}&~~~&\phi \frac{\partial c}{\partial t} 
+ (\bm{v} \cdot \bm{\nabla})c = \bm{\nabla} \cdot (\phi \, \overline{\mathbf{D}} \cdot  \bm{\nabla}c).\label{eq:AD}
\end{eqnarray}
The vectors are noted in bold, with $\bm{v}$ the Darcy velocity vector having $u$ and $w$ as $x$ and $z$-components, and with $\bm{g}$~the gravity acceleration vector pointing in the positive $z$-direction. These equations depend on both the fluid's and porous medium's properties. $P$~is the pressure corrected for the hydrostatic pressure $\rho_0 g z$,  and $\mu\approx 1.15 \times 10^{-3}$~Pa.s is the dynamic viscosity of the salt water considered in this study and at~$20\,^\circ$C~\cite{Sharqawyetal2010,Loodtsetal2014b}. $c$ is the concentration of dissolved CO$_2$ in the fluid normalized by the saturation concentration at the surface given by $C^0_\mathrm{sat}=k_H P^0_{CO_2}$, where $k_H=2.92\times10^{-4}$~mol/m$^3$/Pa is the Henry's constant at the salt concentration and at $20^\circ$C~\cite{Loodtsetal2014b}, and $P^0_{CO_2}$ is the initial CO$_2$ partial pressure above the porous medium. The density difference caused by the CO$_2$ dissolution is given by~$\Delta \rho= \rho_0 \alpha k_H P^0_{CO2}$, where $\alpha$ is the solutal expansion coefficient such as $\alpha k_H=2.38 \times 10^{-9} $\,Pa$^{-1}$ at $20^\circ$C~\cite{Loodtsetal2014b}. Finally, $\overline{\mathbf{D}}$ is an anisotropic diffusion tensor accounting for hydrodynamic dispersion. It is classically expressed as~\cite{HidalgoCarrera2009,EmamiMeybodietal2015,EmamiMeybodi2017b}

\begin{equation}
\label{eq:def_disp_tens}
D_{ij}=\left [ ( D_0+ \alpha_\text{T} \frac{\|\bm{v}\|}{\phi}) \delta_{ij} +(\alpha_\text{L} - \alpha_\text{T} ) \frac{v_{i} v_{j}}{\phi \|\bm{v}\|}  \right ]
\end{equation}
where $D_0=1.24 \times 10^{-9}$ m$^2/$s is the CO$_2$ diffusion coefficient in the salt water solution~\cite{Selletal2013}, $\delta_{ij}$ is the Kronecker symbol, and the notation~$\| \bm{v} \|$ is the Euclidean norm of vector~$\bm{v}$. In the literature, the two length scales~$\alpha_\text{L}$ and $\alpha_\text{T}$ are estimated to be about $\alpha_\text{L} \sim d$ and $\alpha_\text{T}\sim d/10$~\cite{Souzyetal2020,Liangetal2018}, where~$d$ is the typical grain size.

This convective dissolution problem is characterised by the Rayleigh and Darcy numbers~\cite{EnnisKingetal2005,Riazetal2006,Hassanzadehetal2007,EleniusJohannsen2012,Vremeetal2016}
\begin{eqnarray}
\textrm{Ra}&=&\frac{\Delta\rho g K H}{\mu \phi D_0},\label{eq:Ra}\\
\textrm{Da}&=&\frac{K}{H^2}.\label{eq:Da}
\end{eqnarray}
By varying the different experimental parameters (see Table~\ref{tab:param}), and mainly the initial CO$_2$ partial pressure, the Rayleigh number varies from about~$30$ to~$750$. It therefore corresponds to unstable configurations since the critical Rayleigh number~$\textrm{Ra}_c$ above which the instability starts has been found theoretically to be about $32$~\cite{EmamiMeybodietal2015}. With the parameters given above and in Table~\ref{tab:param}, the Darcy number belongs to a limited range between $\textrm{Da}=0.2 \times 10^{-7}$ and $0.5 \times 10^{-7}$. These values are sufficiently small to prevent any Brinkman effects which were shown to occur when $\textrm{Ra}\textrm{Da}^{1/2}$ becomes of order one~\cite{Vremeetal2016}. Here, this parameter remains much smaller than~$1$. As it will be shown later, this value is sufficiently small to get a correct separation of scale by about at least one order of magnitude between the wavelength of the instability and the typical grain size.

\begin{table}[h!]
\caption{Different parameters and dimensionless numbers varied in the experiments with FEP grains. 
\label{tab:param}
}
\begin{ruledtabular}
\begin{tabular}{llllll}
height & thickness & CO$_2$ partial pressure & Rayleigh number & Darcy number & other number \\
\colrule
$H$~[mm] & $e$ [mm] & $P^0_{CO_2}$ [bar] & Ra & Da $\times 10^{7}$ & RaDa$^{1/2}$ \\
$138-220$ & $7$ or $15$ & $0.05-1$ & $30-750$ & $0.2-0.5$ & $0.006-0.12$\\
\end{tabular}
\end{ruledtabular}
\end{table}

\section{Overview of the convective dissolution\label{sec:overview}}

\begin{figure}[t!]
	\begin{centering}
 	\includegraphics[width=1\textwidth]{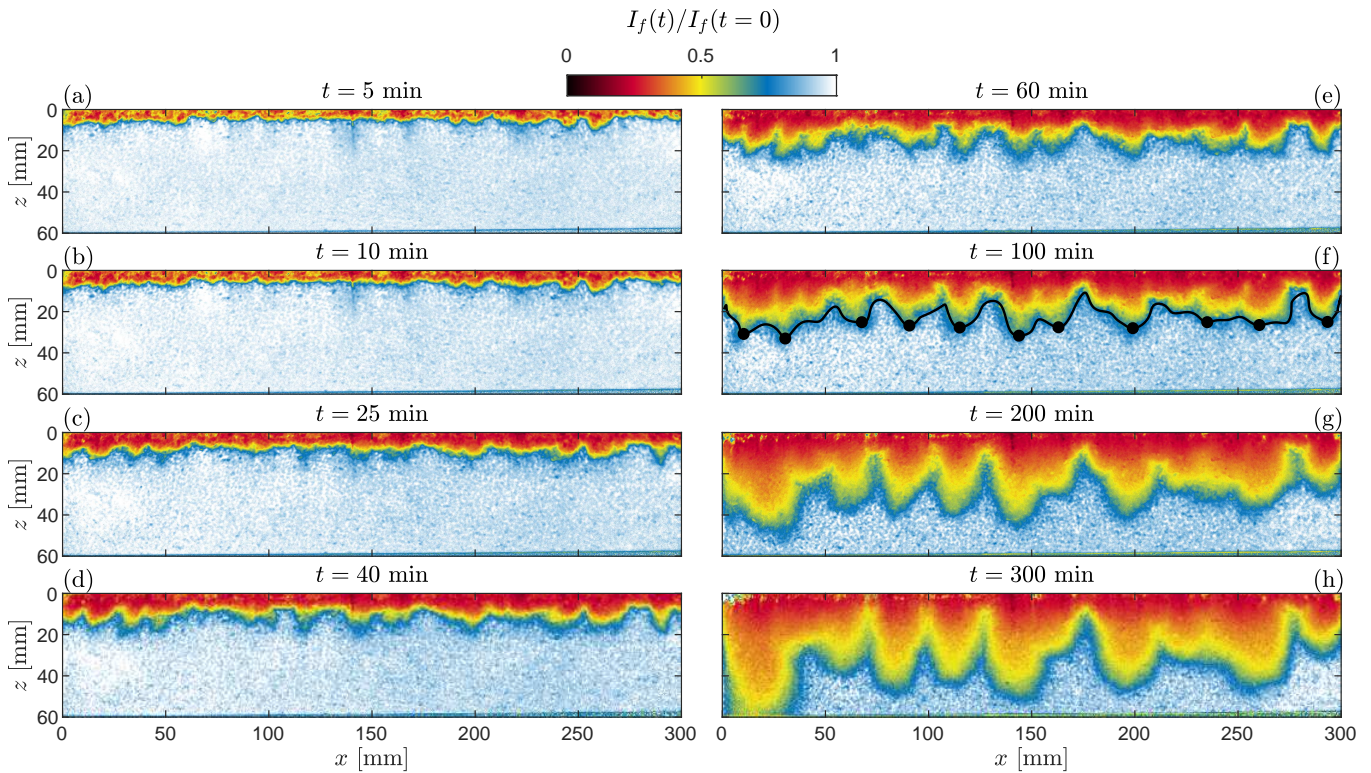}
		\caption{Sequence of images illustrating the instability process. The corresponding times are indicated on top of each image. Panel (f) shows an example of front (solid line) and finger tip (black dots) detection. The initial partial pressure is $P^0_{CO2}=0.6$~bar and the thickness of the tank is $7$~mm. Note that the bottom of each image does not correspond to the bottom of the tank. A complete video corresponding to this image sequence is available in the supplementary materials.\label{fig:image_sequence}}
	\end{centering}
\end{figure}

\begin{figure}[t!]
	\begin{centering}
		\includegraphics[width=0.55\textwidth]{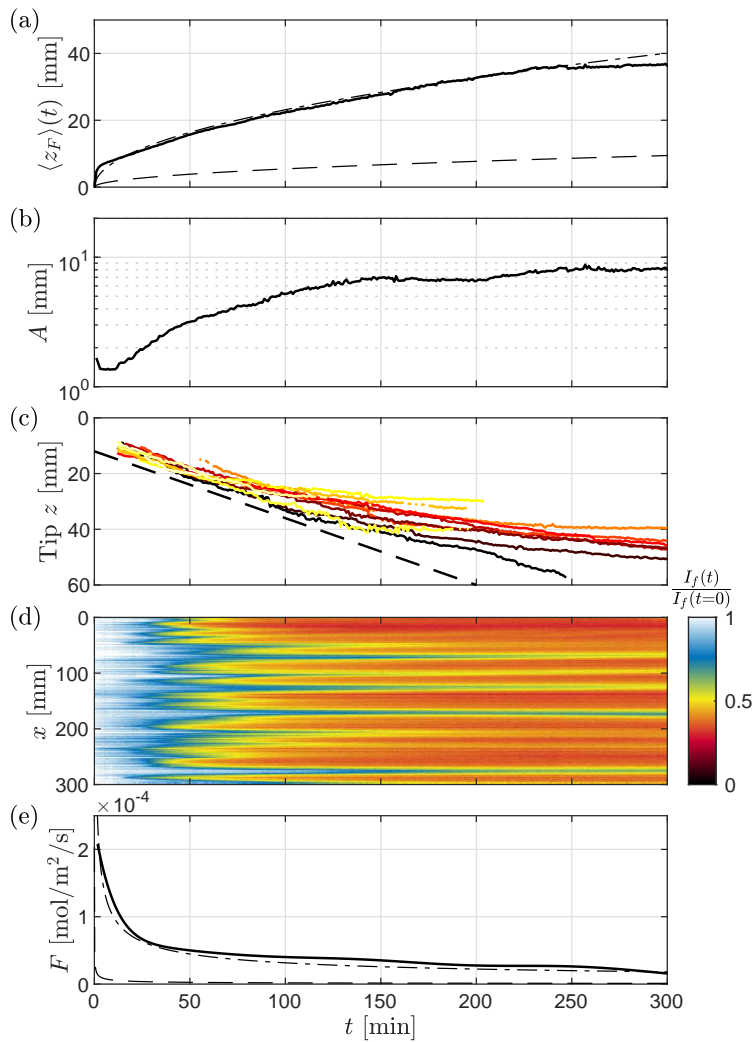}
		\caption{Global measures of the instability. (a)~Mean front depth~$\langle z_F \rangle$ as a function of time. The dashed line represents the expected diffusive front depth $2\sqrt{D_0t}$ while the dashed dotted line highlights a diffusive behavior with a higher diffusion coefficient $D_\text{eff}=18 D_0$. (b)~Corrugation front amplitude~$A$ defined in Eq.~(\ref{eq:front_amp}) as a function of time~$t$ in a log-lin scale. (c)~Fingertip depth (see Fig.~\ref{fig:image_sequence}(f) for an example of fingertip detection) as a function of time. Different colors are used to guide the eyes and easily follow the different fingertips. The dashed line shows a constant velocity of about $4 \times 10^{-3}$~mm/s for the fingertips at the beginning of the experiment. (d)~Intensity profile at $z=15$~mm as a function of time. White-blue regions correspond to low CO$_2$ concentrations and red/dark regions to high CO$_2$ concentration. (e)~Temporal evolution of the CO$_2$ flux measured at the interface (solid line). The theoretical diffusive flux defined by Eq.~\eqref{eq:DiffusiveFlux} is plotted as a dashed line while the diffusive flux with a higher diffusion coefficient $D_\text{eff}=290 D_0$ is shown by a dashed dotted line. The initial CO$_2$ partial pressure is $P^0_{CO2}=0.6$~bar and the thickness of the tank is $7$~mm.\label{fig:front_flux_fingers}}
	\end{centering}
\end{figure}

Figure~\ref{fig:image_sequence} presents the temporal evolution of the convective dissolution process for an intermediate Rayleigh number ($\textrm{Ra}=513$), in the top region of the porous medium ($z=0-60$~mm). The visualisation initially exhibits a thin layer with warm colors at the free surface, corresponding to CO$_2$-rich brine where the pH is reduced, thus lowering the emitted intensity. This layer becomes thicker as time evolves and fingers start to develop (Figs.~\ref{fig:front_flux_fingers}(a), (b) and (c)), grow with time (Figs.~\ref{fig:front_flux_fingers}(d), (e) and (f)) and merge (Figs.~\ref{fig:front_flux_fingers}(f), (g) and (h)). At first glance, these visualisations are very similar to the ones obtained in Hele-Shaw cells without grains~\cite{Vremeetal2016}. However, the fingers seem to be smoother and broader, as the separation between the fingers is not as clear as in Hele-Shaw experiments. This is likely to be due to hydrodynamic dispersion, which enhances the mixing between CO$_2$-rich brine and pure brine as compared to pure molecular diffusion~\cite{Wangetal2016,Liangetal2018}.

Although the relation between the intensity and the CO$_2$ concentration is not completely quantitative~\cite{Thomasetal2015}, these visualisations can be used to determine the depth of the front at a given intensity. For example the front $z_F(x,t)$ is plotted at a given time~$t$ in Fig.~\ref{fig:image_sequence}(f). From this front, it is easy to define at each time~$t$ the depth of the fingertips $z_i(t)$ corresponding to local maxima of the function $z_F(x,t)$. These fingertips are marked by dots in Fig.~\ref{fig:image_sequence}(f). They will be used in the following to estimate qualitatively the wavelength of the instability and to characterize quantitatively the velocity of the fingers.

At each time~$t$, the front $z_F(x,t)$ is also used to measure the front corrugation amplitude~$A(t)$, which is defined as the standard deviation of the front~$z_F(x,t)$ along~$x$~\cite{Vremeetal2016}:
\begin{equation}
A(t)=\langle z_F^2(x,t)-\langle z_F \rangle^2(t)\rangle^{1/2},\label{eq:front_amp}
\end{equation}
where $\langle \rangle$ denotes the average over~$x$. The mean front depth~$\langle z_F \rangle$ and the front corrugation amplitude~$A$ are plotted in Figs.~\ref{fig:front_flux_fingers}(a) and~(b) as a function of time, with a classical log-lin scale for the front corrugation amplitude. As expected, the mean front depth~$\langle z_F \rangle$ increases as a function of time, showing that the CO$_2$-rich brine penetrates into the porous medium. The time behavior of the mean front depth~$\langle z_F \rangle$ is compatible with a global diffusive process in~$\sqrt{t}$ (shown by the dashed dotted line in Fig.~\ref{fig:front_flux_fingers}(a)), despite the fact that convection clearly happens at short time scales according to Fig.~\ref{fig:image_sequence}. Indeed, the thickness of the diffusive layer given by $2\sqrt{D_0t}$ is not in agreement when using the CO$_2$ diffusion coefficient $D_0$, as shown by the dashed line in Fig.~\ref{fig:front_flux_fingers}(a): the dashed dotted line that matches the mean front depth is represented for an effective diffusion coefficient~$D_\text{eff}=18D_0$. In Fig.~\ref{fig:front_flux_fingers}(b), the front corrugation amplitude~$A$ exhibits as a function of time a small initial decay followed by an important growth after 10 minutes. This is classical of the linear instability obtained in Hele-Shaw cells without grains~\cite{Vremeetal2016} or in numerical simulations at the Darcy scale~\cite{EleniusJohannsen2012,Slimetal2013,Slim2014}. However, it is hard to define a clear exponential growth of the amplitude in these experiments since the curve departs quickly from the straight line (after about 40 minutes) but keeps increasing slowly until at least $300$~minutes. A detailed quantitative characterization of this corrugation growth will be given in section~\ref{instability_charac}.

The depth of the different fingertips is shown in Fig.~\ref{fig:front_flux_fingers}(c) as a function of time. At the beginning of the experiment, the fingertips plunge with a constant velocity, as suggested by the dashed line. Some of these fingertips disappear due to the merging of the plumes. This is also characteristic of the \textit{flux-growth} and \textit{constant-flux regimes} of the instability~\cite{Slim2014}, that occur after the \textit{diffusive regime}. However, it seems that the merging events are less frequent here than in classical Hele-Shaw experiments or numerical simulations~\cite{Vremeetal2016,Slimetal2013,Slim2014}. This is also visible in Fig.~\ref{fig:front_flux_fingers}(d), which exhibits horizontal intensity profiles at $z=15$~mm as a function of time. As in Fig.~\ref{fig:image_sequence}, the fingers appear much broader and seem to occupy a larger portion of the width than in previous experimental studies without grains. These features therefore arise from the presence of the grains, and are likely due to hydrodynamic dispersion~\cite{Wangetal2016,Liangetal2018}.

Figure~\ref{fig:front_flux_fingers}(e) shows the CO$_2$ flux as a function of time, measured from the decay of the CO$_2$ partial pressure. The flux is large at the beginning and globally decreases as a function of time. This typical behavior is also classical, as it has been observed in several experiments in PVT cells~\cite{KneafseyPruess2011,NazariMoghaddametal2012,Seyyedietal2014,NazariMoghaddametal2015}. The flux is expected to be diffusive at early stages, before being enhanced by the start of the convection as observed in numerical simulations obtained at the Darcy scale~\cite{EleniusJohannsen2012,Slim2014,Tilton2018} or experiments in a Hele-Shaw cell~\cite{Slimetal2013}. However, for the experiment presented in Fig.~\ref{fig:front_flux_fingers}, no enhancement of the flux is visible. In addition, the flux obtained here is about one order of magnitude larger than the diffusive flux given by~\cite{Slimetal2013,Slim2014,DePaolietal2017}
\begin{equation} 
F_{\mathrm{diff}}=\phi k_H P^0_{CO_2}\sqrt{\frac{D_0}{\pi t}} \label{eq:DiffusiveFlux}
\end{equation}  
where the dispersion tensor is reduced to molecular diffusion in the absence of flow. Eq.~\eqref{eq:DiffusiveFlux} is plotted as a dashed line in Fig.~\ref{fig:front_flux_fingers}(e). Again, the time behavior seems compatible with a diffusive phenomenon, as shown by the dashed dotted line in Fig.~\ref{fig:front_flux_fingers}(e), but with an effective diffusion coefficient~$D_\text{eff}=290D_0$. Note that this effective value of the diffusion coefficient is not incompatible with the one given for the mean front depth~$\langle z_F \rangle$ in Fig.~\ref{fig:front_flux_fingers}(a), as the flux also depends on the fingering pattern and not only on the mean front depth. The dynamics of the flux as a function of the initial CO$_2$ partial pressure will be described in more detail in section~\ref{sec:Flux}.

\section{Instability characteristics\label{instability_charac}}

\subsection{Measured and predicted growth rate}

\begin{figure}[t!]
	\begin{centering}
		\includegraphics[width=0.75\textwidth]{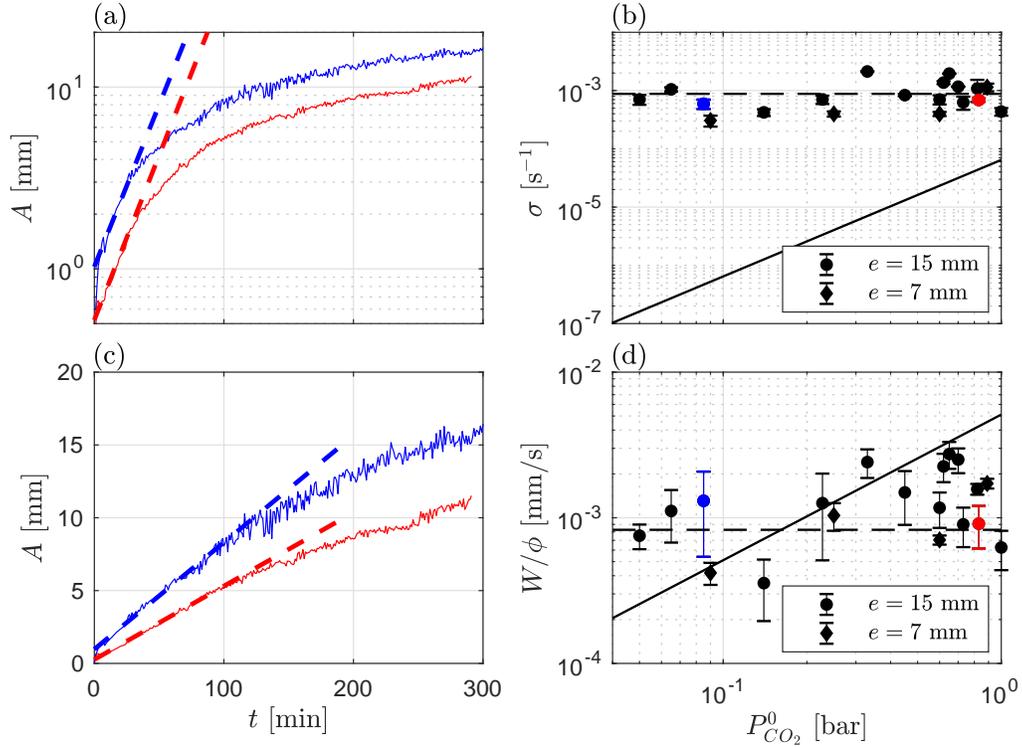}
		\caption{Analysis of the front corrugation amplitude growth. (a)~Front corrugation amplitude~$A$ as a function of time~$t$ in a log-lin scale for two different experiments, at low (blue) and high (red) CO$_2$ partial pressures. The thick dashed lines correspond to the fits by an exponential function to determine the growth rate. (b)~Growth rate~$\sigma$ as a function of initial CO$_2$ partial pressure in log-log scales. The solid line represents the theoretical prediction~\cite{EleniusJohannsen2012} while the dashed line shows the trend obtained in the experiments. (c)~Front corrugation amplitude~$A$ as a function of time~$t$ in a lin-lin scale for the same two experiments shown in panel (a). The thick dashed lines correspond to the fits by a linear function to determine the interstitial forcing velocity~$W/\phi$ (see Eq.~(\ref{eq:amp_linear})). (d)~Interstitial forcing velocity~$W/\phi$ (see Section~\ref{sec:forcing}) as a function of initial CO$_2$ partial pressure in log-log scales. The dashed line represents the trend observed in the experiments, while the solid one corresponds to the typical interstitial velocity~$\mathcal{U}/\phi$. The experiments exhibited in panels (a) and (c) are marked with corresponding colors in panels (b) and (d).\label{fig:front_characteristics}}
	\end{centering}
\end{figure}

The growth rate of the linear instability is classically measured by plotting the front corrugation amplitude~$A$ as a function of time in a log-lin plot~\cite{Slim2014,Vremeetal2016}, as in Fig.~\ref{fig:front_flux_fingers}(b) or in Fig.~\ref{fig:front_characteristics}(a). The growth rate is then determined by fitting the curve with a line, as shown in Fig.~\ref{fig:front_characteristics}(a). By performing this fitting operation on all the experiments performed with the FEP grains, we can extract the growth rate~$\sigma$ and plot it as a function of the initial CO$_2$ partial pressure, the main quantity that can be varied experimentally. Fig.~\ref{fig:front_characteristics}(b) shows the plot, for the two tank thicknesses considered here. First, the results are independent of the tank thickness. This tends to indicate that the flow at the Darcy scale is 2-D (i.e. independent of $y$). However, more surprisingly, the growth rate seems to also be independent of the initial CO$_2$ partial pressure. Indeed, it remains close to a mean value equal to $\sigma \approx 0.9\times 10^{-3}$~s$^{-1}$, represented by the dashed dotted line in Fig.~\ref{fig:front_characteristics}(b). The error bars have been obtained by varying the intensity threshold for the front detection, thus slightly changing the value obtained by the exponential fit of the front corrugation amplitude~$A$. 

At the Darcy scale, the three governing equations~(\ref{eq:incomp})-(\ref{eq:AD}) can be made dimensionless using the following characteristic quantities~\cite{EleniusJohannsen2012,Slimetal2013,Slim2014,EmamiMeybodietal2015,DePaolietal2017}:
\begin{eqnarray}
\mathcal{U}&\equiv&\frac{K \Delta \rho g}{\mu}~~~\textrm{for fluid velocities},\\
\mathcal{L}&\equiv&\frac{\phi D_0}{\mathcal{U}}=\frac{\mu \phi D_0}{K \Delta \rho g}~~~\textrm{for length scales},\label{eq:length}\\
\mathcal{T}&\equiv&\frac{\phi \mathcal{L}}{\mathcal{U}}=\frac{(\mu \phi)^2 D_0}{(K \Delta \rho g)^2}~~~\textrm{for time scales}.\label{eq:time}
\end{eqnarray}
In that case, the dimensionless equations become independent of the Rayleigh and Darcy numbers. Indeed, the Rayleigh number simply imposes the boundary condition at the bottom of the reservoir~\cite{EleniusJohannsen2012}, which has no influence on the onset of the instability as it takes place close to the surface. As a consequence, the growth rate is independent of the Rayleigh and Darcy number in this formulation and was found to be equal to $\sigma^\star=3 \times 10^{-3}$ numerically~\cite{EleniusJohannsen2012}, and further confirmed experimentally in a Hele-Shaw cell~\cite{Vremeetal2016}. Note, however, that this growth rate has been obtained without accounting for hydrodynamic dispersion, i.e. by simply assuming~$\overline{\mathbf{D}}=D_0 \overline{\mathbf{I}}$ in Eq.~(\ref{eq:def_disp_tens}), with $\overline{\mathbf{I}}$ the identity tensor. Going back to dimensional units, the theoretical prediction of the growth rate is therefore simply given by
\begin{equation}
\sigma = \frac{\sigma^\star}{\mathcal{T}} = 3 \times 10^{-3} \frac{(K \rho_0 \alpha k_H P_{CO_2}^0 g)^2}{(\mu \Phi)^2 D_0}.\label{eq:GrowthRateTh}
\end{equation}

This theoretical prediction is plotted as a solid line in Fig.~\ref{fig:front_characteristics}(b). It is clear that the theory underestimates the growth rate by about one to three orders of magnitude, depending on the CO$_2$ partial pressure. This is actually a consequence of the scaling of the theoretical growth rate as $(P_{CO_2}^0)^2$ in Eq.~(\ref{eq:GrowthRateTh}) whereas the experimental growth rate does not seem to depend on $P^0_{CO_2}$. This discrepancy clearly indicates that the growth of the front corrugation amplitude in our experiments is due to a different mechanism, that may be related to the pore scale flow heterogeneity induced by the presence of the grains. By looking for a relationship between the measured growth rate and the grains, one can remark that the value of the measured growth rate~$\sigma$ is of the order of $D_0/d^2$, with $d=1.5$~mm taken as the mean dimension of the grains.

\subsection{Forcing by porosity fluctuations\label{sec:forcing}}

To explain our experimental observations, we here propose a model based on a forcing of the instability by porosity fluctuations. Such a mechanism has already been discussed theoretically in a recent paper by Tilton~\cite{Tilton2018}. In this work, he has considered sinusoidal porosity fluctuations and shown that they trigger the instability earlier. Note that this work has been done without considering hydrodynamic dispersion, as it drastically makes the calculation more complex. Here, we reproduce his calculations using dimensional quantities. We now consider porosity fluctuations such that
\begin{equation}
\phi = \langle \phi \rangle [1 + \varepsilon \widetilde\phi (x,z) ].
\end{equation}
Here, $\langle \phi \rangle$ is the spatially averaged porosity while $\widetilde\phi$ stands for porosity fluctuations and $\varepsilon$ is assumed to be small. It is important to note that the porosity variations are considered over length scales larger than the typical grain size~$d$, as Eqs.~(\ref{eq:incomp}),~(\ref{eq:Darcy}) and~(\ref{eq:AD}) are defined by averaging pore-scale continuity, flow and transport equations~\cite{Tilton2018}. Following Tilton~\cite{Tilton2018}, we then perform an asymptotic expansion of Eqs.~(\ref{eq:incomp}),~(\ref{eq:Darcy}) and~(\ref{eq:AD}) by seeking a solution of the form
\begin{equation}
c(x,z,t) \approx c_b(z,t) + \varepsilon c_1(x,z,t) + ...,
\end{equation}
with similar expansions for the horizontal and vertical velocities~$u$ and~$w$. At zero-order ($\varepsilon^0$), there is no perturbation and the zero-order solution describes the CO$_2$ diffusion at the top of the porous medium
\begin{equation}
\bm{v_b}=\bm{0}~~~\textrm{and}~~~c_b(z,t)=1-\mathrm{erf}\left(\frac{z}{2\sqrt{D_0t}}\right)\equiv \frac{2}{\sqrt{\pi}}\int_{z/(2\sqrt{D_0t})}^{+\infty} \exp(-\tau^2)\textrm{d}\tau.\label{eq:diff}
\end{equation}
As a consequence, the order~$\varepsilon^1$ represents the typical velocities created by the instability. It is governed by the following equations~\cite{Tilton2018}:
\begin{equation}
\frac{\partial u_1}{\partial x} + \frac{\partial w_1}{\partial z}=0,~~~~~\bm\nabla^2 w_1= \frac{K \Delta \rho g}{\mu}\frac{\partial^2 c_1}{\partial x^2},~~~~~\langle \phi \rangle \frac{\partial c_1}{\partial t} + w_1 \frac{\partial c_b}{\partial z} - \langle \phi \rangle D_0 \bm\nabla^2 c_1 = F_1,\label{eq:order1}
\end{equation}
where $F_1$ is a forcing term that can be reduced to
\begin{equation}
F_1=\langle \phi \rangle D_0 \frac{\partial \widetilde \phi}{\partial z}\frac{\partial c_b}{\partial z}.
\end{equation}
It therefore appears that the order~$\varepsilon^1$ of the asymptotic expansion is forced by the vertical variation of porosity multiplied by the vertical gradient of the diffusion concentration field~$c_b$. Note that this term is strictly null when one does not consider porosity fluctuations, as in classical numerical simulations using Darcy's law with a uniform porosity~\cite{EleniusJohannsen2012,Slim2014} or Hele-Shaw experiments mimicking experimentally a 2-D porous medium of uniform porosity and permeability~\cite{Vremeetal2016,Slimetal2013}. However, in our case, we investigate a 3-D granular porous medium, which is by nature random below the representative elementary volume (REV) characteristic of the Darcy scale~\cite{deMarsilly2004}. One can therefore expect to have porosity fluctuations that may force some initial velocities. Note however that this implies to some extent that the spatial averaging from pore to Darcy scale is done over a scale that is not quite sufficiently large to be a REV; in such a way pore scale fluctuations are accounted for in the Darcy scale description as fluctuations in the porosity field, though the porous medium would probably be homogeneous at the Darcy scale if the latter were defined by averaging over a REV.

At the diffusive front, the vertical concentration gradient is initially very large since it scales as $1/\sqrt{D_0t}$ (as given by the $z$-derivative of the second equation of~(\ref{eq:diff})). At early stages, the two dominant terms in the last equation of~\eqref{eq:order1} are thus $w_1 (\partial c_b/\partial z) $ and $F_1$. Equating these two terms leads to the approximation: 
\begin{equation}
 w_1 \approx \langle \phi \rangle D_0 \frac{\partial \widetilde \phi}{\partial z}.
\end{equation}
As a consequence, the vertical velocity at order $\varepsilon^1$ is directly forced by the vertical variations of porosity fluctuations. This leads to a forcing velocity
\begin{equation}
W=\varepsilon w_1 = D _0 \frac{\partial \phi}{\partial z},
\end{equation}
which is expressed as a Darcy velocity, as it is derived from Eqs.~(\ref{eq:incomp})-(\ref{eq:AD}). Note that these vertical variations of porosity may occur at several scales, somehow correlated with the typical grain size~$d$.

One can now estimate the evolution of the front corrugation amplitude~$A$. Indeed, the front $z_F$ is advected by the interstitial (i.e. mean pore scale) vertical forcing velocity~$W/\phi$, such that the front corrugation amplitude $A$ is expected to increase linearly in time in the early stage. This is actually what is observed in the experiments. Indeed, plotting the front corrugation amplitude in linear scale (as done in Fig.~\ref{fig:front_characteristics}(c)) indicates that $A$ increases linearly  during about 100 minutes. This linear fitting is more efficient than the exponential fitting previously done to extract the growth rate (which seemed to be reasonable only for 40 minutes). The slope of the linear growth of $A$, i.e. the interstitial forcing velocity $W/\phi$, has been measured for all the experiments performed with the FEP grains, and error bars have been obtained by varying the intensity threshold for the front detection. The interstitial forcing velocity~$W/\phi$ is plotted as a function of the initial CO$_2$ partial pressure in Fig.~\ref{fig:front_characteristics}(d), and appears independent of this parameter as shown by the dashed line, with a mean value $W/\phi\approx 10^{-3}$~mm/s. This value is fully compatible with the value $4  \times 10^{-3}$ mm/s  obtained by following the finger tip depth as a function of time, in Fig.~\ref{fig:front_flux_fingers}(c). Note that the dispersion of the value found for~$W/\phi$ is larger than the one found for~$\sigma$. Indeed, the linear fit is more sensitive to the value of the iso-concentration than the exponential fit. However, Fig.~\ref{fig:front_characteristics}(d) shows the relative dispersion of data, which remains limited. 
For comparison, the typical interstitial velocity~$\mathcal{U}/\phi$, proportional to~$P^0_{CO_2}$, is plotted with a solid line in Fig.~\ref{fig:front_characteristics}(d). Both velocities remain on the same order of magnitude, but the scaling with~$P^0_{CO_2}$ is clearly different. The experimental results on the forcing velocity $W/\phi$ are therefore compatible with the value $D_0 (\partial \phi/\partial z) /\phi$, which is independent of~$P^0_{CO_2}$. Moreover, if one assumes that~$(\partial \phi/\partial z) \sim \phi/d$, one can recover quantitatively the mean value of~$W/\phi$ measured in the experiments. This suggests a strong connection between porosity fluctuations and the typical grain size~$d$, as expected since these fluctuations are the Darcy scale signature of pore scale heterogeneity. Note, however, that in a classic Darcy approach, local averaging is performed over a REV sufficiently large for porosity fluctuations to be negligible in the Darcy description of such a granular porous medium, whose grain size distribution is peaked and not too wide. Here Tilton's model then implies that the REV size is chosen small enough for pore scale heterogeneity to translate into spatial fluctuations of the porosity field.

At late times, the concentration gradient $\partial c_b/\partial z$ eventually becomes small enough such that the forcing term may become negligible. In that case, the equations become the classical equations in a homogeneous porous medium. The front corrugation amplitude $A$ is thus expected to grow exponentially with a growth rate $\sigma$ given by Eq.~\eqref{eq:GrowthRateTh}. A hybrid equation for~$A$ can be given as
\begin{equation}
\frac{\textrm{d}A}{\textrm{d}t}=\sigma A + \frac{D_0}{\phi}\frac{\partial \phi}{\partial z},\label{eq:front_amp_eq}
\end{equation}
where the first term is the growth of the instability as given by the theoretical predictions~\cite{EleniusJohannsen2012} and the second term is the forcing term acting on the velocity field, so on the time derivative of the front corrugation amplitude~$A$. The solution of Eq.~(\ref{eq:front_amp_eq}) is given by
\begin{equation}
A(t)= \frac{D_0}{\phi} \frac{\partial \phi}{\partial z}\left(\frac{\exp(\sigma t)-1}{\sigma}\right) \approx \frac{D_0}{\phi}\frac{\partial \phi}{\partial z} t~~~~~\textrm{for}~\sigma t \ll 1.\label{eq:amp_linear}
\end{equation}

In our experiments, the initial growth of the front amplitude occurs generally during about $100$~min, while the growth rate is expected to vary between $10^{-7}$ and $10^{-4}$~s$^{-1}$ in the pressure range investigated (see Fig~\ref{fig:front_characteristics}(b)). Therefore, in this work, we are always in the limit where $\sigma t$ is smaller than, or of order, ~$1$. This means that the linear fit for the front amplitude is always applicable here. For higher CO$_2$ partial pressure, one can expect to have $\sigma t \gg 1$, so a clear exponential (and not a linear) growth. However, the tank used here does not allow us to go higher than $P^0_{CO_2}=1$~bar.

\subsection{Wavelength}

After the growth rate, we now focus on the wavelength of the instability that has also been significantly discussed in the literature. Indeed, as the grains seems to have a strong influence on the growth rate with the forcing by porosity fluctuations, it is worth investigating the effect of the grains on this other important parameter of the convective dissolution instability. Note that the typical convective length scale~$\mathcal{L}$ (see Eq.~(\ref{eq:length})) belongs to the range~$[0.1-1.9]$~mm, which means that the grains are typically larger than, or about the same size as, this typical scale. However, as stated further in this section, the front wavelength~$\lambda$ is theoretically expected to be about two orders of magnitude larger than the typical convective length scale~$\mathcal{L}$~\cite{Riazetal2006,Hassanzadehetal2007}, so much larger than the typical grain size. This is further confirmed by the experimental results, meaning that there is a clear scale separation in the experiments.

\begin{figure}[t!]
	\begin{centering}
		\includegraphics[width=1\textwidth]{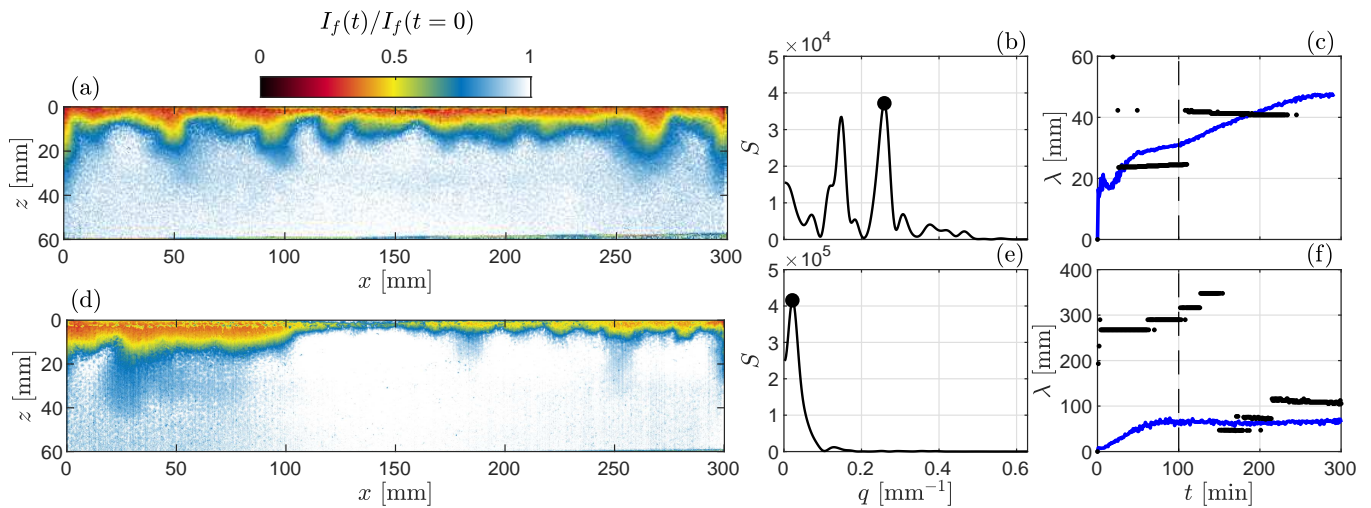}
		\caption{Analysis of the front's dominant wavelength. (a)~Snapshot at $t=100$~min for an experiment at high CO$_2$ partial pressure ($P^0_{CO_2}=0.83$~bar), shown in red in Fig.~\ref{fig:front_characteristics}. (b)~Spectrum~$S$ of the front for the image shown in panel (a). The maximum of the spectrum is indicated by a black dot. (c) Wavelength~$\lambda$ as a function of time determined using two different methods: the Fourier transform of the front (black dots)~\cite{Vremeetal2016} and Eq.~(\ref{eq:Slim}) (blue line)~\cite{Slim2014}. The vertical dashed line indicates the time of the snapshot shown in panel (a). (d), (e) and (f):~Same panels as (a), (b) and (c) but for an experiment at low CO$_2$ partial pressure ($P^0_{CO_2}=0.085$~bar), shown in blue in Fig.~\ref{fig:front_characteristics}.\label{fig:lambda}}
	\end{centering}
\end{figure}

Depending on the initial CO$_2$ partial pressure, the experiments qualitatively exhibit different length scales for the fingers. This is illustrated in Figs.~\ref{fig:lambda}(a) and (d), where panel~(a) corresponds to a high CO$_2$ partial pressure and panel~(d) to a low CO$_2$ partial pressure. In Fig.~\ref{fig:lambda}(a), one can clearly see about 12 fingertips within the full width of the tank, while in Fig.~\ref{fig:lambda}(c), one mainly observes a large fingertip on the left of the tank in addition to smaller fingertips disturbing this large scale feature locally (particularly to the right of the tank). This illustrates well that the front exhibits a multi-scale nature.

In order to experimentally obtain a typical wavelength for each experiment, different methods have been described in the literature,  two of which have been used on the experimental data. The first method estimates the wavelength from the maximum of the Fourier spectrum of the front~$z_F(x,t)$ at a given time~$t$~\cite{Vremeetal2016}. Figures~\ref{fig:lambda}(b) and~(e) exhibit examples of the spectra obtained from the fronts detected in Fig.~\ref{fig:lambda}(a) and~(d). The maximum of each spectrum is marked with a black dot but the spectrum at high CO$_2$ partial pressure clearly exhibits additional peaks. This also highlights the multi-scale nature of the front. The second method defines the typical finger width as $\pi/q$ and therefore the wavelength as $\lambda=2\pi/q$ where the horizontal wavenumber~$q$ is given by~\cite{Slim2014}
\begin{equation}
q=\sqrt{\frac{\langle(\partial c'/\partial x)^2\rangle}{\langle c'^2 \rangle}},\label{eq:Slim}
\end{equation}
with $c'=c-\langle c \rangle$ is the concentration fluctuations and $\langle \rangle$ represents the average in the horizontal direction~$x$. Note that this relation is exact when $c'$ is a sinusoidal function of the variable~$qx$. In our case, the concentration~$c$ has been replaced by the normalised intensity~$I_f(x,z,t)/I_f(x,z,t=0)$ obtained using the planar laser induced fluorescence technique (see Section~\ref{PLIF}) and taken at~$z=\langle z_F \rangle(t)$.

These two different methods can be applied for each image, in order to follow the evolution of the estimated wavelength with time. This  evolution is shown in Figs.~\ref{fig:lambda}(c) and (f) for both methods described previously. At high CO$_2$ partial pressure (see Fig.~\ref{fig:lambda}(c)), despite experimental noise and once the instability has started, both methods give the same value of the wavelength within about $50\,\%$ (a few centimetres), in qualitative agreement with the snapshot of the instability in Fig.~\ref{fig:lambda}(a). This wavelength slightly increases with time. This is expected as the different fingers initially produced by the instability merge with time, thus increasing the typical wavelength with time. This merging process is visible in Fig.~\ref{fig:image_sequence}. However, for low CO$_2$ partial pressures, the two methods result in different wavelengths in Fig.~\ref{fig:lambda}(f). Indeed, the second method based on the finger width gives a wavelength on the order of about $70$~mm while the first one shows a typical wavelength of about~$300$~mm corresponding to the tank width until~$t=150$~min. As stated before, one large finger can be identified on the left side of the front in Fig.~\ref{fig:lambda}(d). The wavelength given by the spectrum finally converges toward a value close to the one obtained by the second method at the end of the experiment, when the small fingers on the right side of the front have grown sufficiently. The difference between the two methods can be understood since the second method corresponds in fact to the second moment of the spectrum $\int q^2 S(q) \textrm{d}q~/ \int S(q) \textrm{d}q$, which gives more weight to large wavenumbers than the first method. 

\begin{figure}[t!]
	\begin{centering}
		\includegraphics[width=0.8\textwidth]{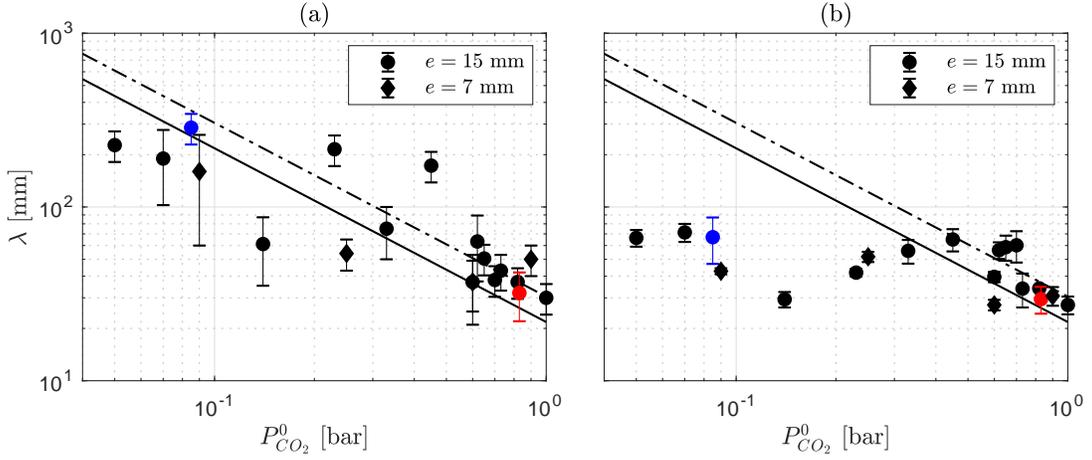}
		\caption{Front wavelength~$\lambda$ as a function of the initial CO$_2$ partial pressure obtained using two different methods, based on (a) the spectrum of the front~$z_F(x,t)$~and (b) the typical finger width (see Eq.~(\ref{eq:Slim})). The solid and dashed dotted lines respectively indicate the theoretical predictions from~Riaz \textit{et al.}~\cite{Riazetal2006} and Hassanzadeh~\textit{et~al.}~\cite{Hassanzadehetal2007}. The experiments exhibited in Fig.~\ref{fig:lambda} are marked with blue and red dots. \label{fig:lambda_result}}
	\end{centering}
\end{figure}

In the literature, the typical time at which the wavelength is measured corresponds to the non-linear time scale~$t_\sigma$~\cite{Vremeetal2016}, i.e. when the front amplitude deviates from the exponential fit. However, in our case, as we have shown previously that a linear fit would be better for the growth of the front amplitude, it is possible to consider another time scale, which is the one where the front amplitude deviates from its initial linear growth, noted~$t_W$. This time depends on the different experiments but remains of order $100-150$~minutes, as shown by the two examples in Fig.~\ref{fig:front_characteristics}(c). We have chosen to measure the wavelength by averaging the values obtained at $5$ different times linearly chosen between~$t_\sigma$ and~$t_W$. This time range corresponds to the end of the linear growth, as the non-linear time scale~$t_\sigma$ is about~$40$ to~$100$~minutes in the different experiments. The error bars obtained for the wavelength measured using the second method correspond to the standard deviation of the $5$~values obtained. For the first method, we have chosen to set the error bars at~$20\,\%$ of the wavelength value, as the estimation of the wavelength by this method remains limited by the multi-scale nature of the front. The results obtained with both methods are respectively shown in Fig.~\ref{fig:lambda_result}(a) and~(b) for the different experiments, i.e. as a function of the initial CO$_2$ partial pressure.

First, similarly to the growth rate, the results are independent of the thickness of the tank, which is in agreement with a 2-D instability at the Darcy scale. In addition, the measured wavelength is not too far from the value predicted by earlier numerical simulations and theoretical derivations at the Darcy scale. Indeed, the dimensionless wavelength (i.e. dimensionalized by $\mathcal{L}=(\mu \phi D_0)/(K \Delta \rho g$)) was found to be equal to $2\pi/0.07\approx90$~\cite{Riazetal2006} or to $40\pi\approx126$~\cite{Hassanzadehetal2007}. In Fig.~\ref{fig:lambda_result}(a), this prediction is in good agreement with the experimental results for high CO$_2$ partial pressures but it slightly overestimates the wavelength by a factor smaller than~$2$ at low CO$_2$ partial pressures. In Fig.~\ref{fig:lambda_result}(b), this prediction still overestimates the results at low CO$_2$ partial pressures, but with a larger factor ($3$ to~$4$), while the agreement is still fair at high CO$_2$ partial pressures. In fact the wavelength obtained with the second method appears rather independent of the CO$_2$ partial pressure and may correspond to the smallest detectable wavelength in the front. However, the overall value of the wavelength determined by both methods is here on the same order of magnitude as the predictions, while this is clearly not the case for the growth rate. These two results may therefore appear in contradiction.

\begin{figure}[t!]
	\begin{centering}
		\includegraphics[width=1\textwidth]{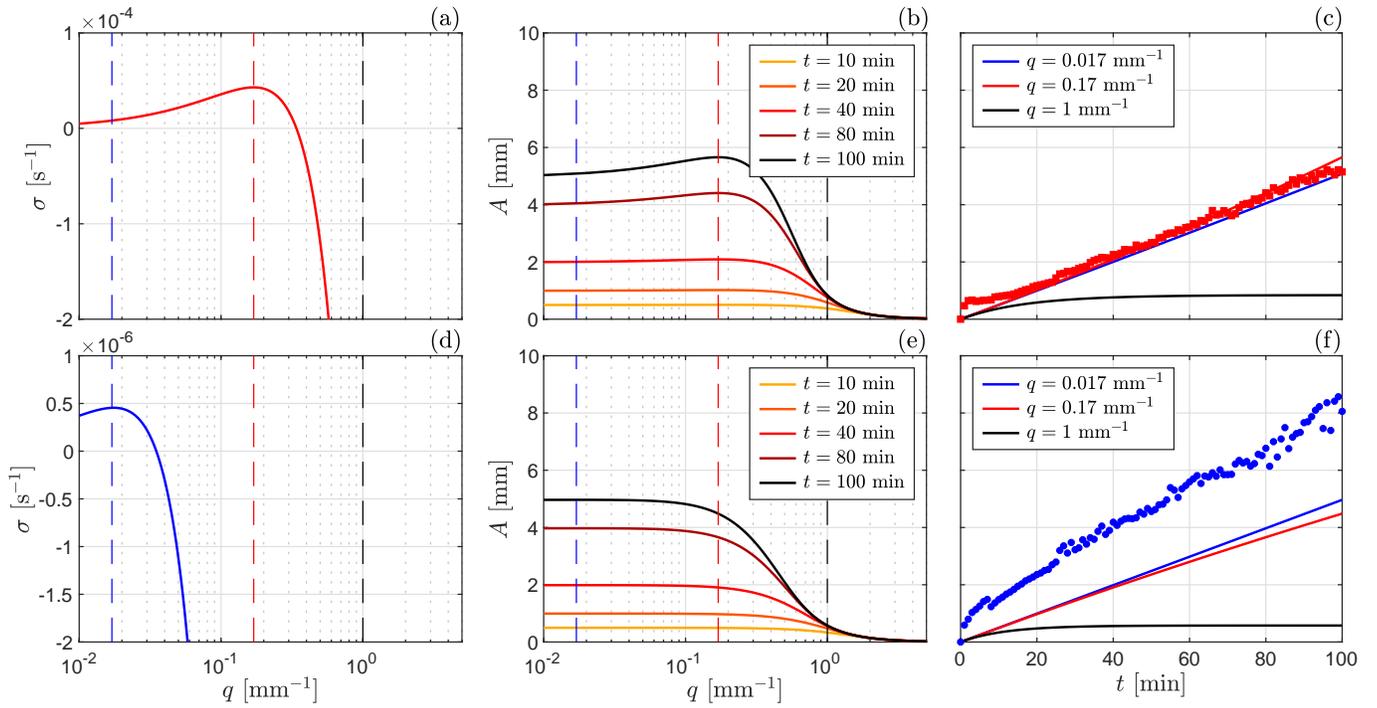}
		\caption{Heuristic model for the growth of the front corrugation amplitude~$A$ as a function of time and wavenumber. The top line corresponds to an experiment at high CO$_2$ partial pressure ($P^0_{CO_2}=0.83$~bar) while the bottom one is for an experiment at low CO$_2$ partial pressure ($P^0_{CO_2}=0.085$~bar). These experiments are the same as the ones addressed in Figs.~\ref{fig:front_characteristics} and~\ref{fig:lambda}. The left column illustrates the dependence of the growth rate~$\sigma$ as a function of the wavenumber~$q$ using the heuristic model~(\ref{eq:heuristic}). The central and right columns show the evolution of the front corrugation amplitude~$A$ (given by Eq.~\eqref{eq:amp_linear} with $\sigma(q)$ set by Eq.~\eqref{eq:heuristic}) as a function of the horizontal wavenumber and time respectively. The colors of the vertical dashed lines marking different wavenumbers in the left and central columns  correspond to the ones used in the right column. The experimental data, also shown in Fig.~\ref{fig:front_characteristics}(c), are represented in panels~(c) (red squares) and~(f) (blue dots) as a reference.
	\label{fig:spectrum}}
	\end{centering}
\end{figure}

Nevertheless, the forcing mechanism due to porosity fluctuations is assumed to force with a broad spectrum in the horizontal direction. It is thus possible that the instability still selects the most unstable wavelength although the growth of the fingers is much faster than in the absence of heterogeneities. For example, taking the model \eqref{eq:amp_linear} developed for the front corrugation amplitude~$A$ with a heuristic growth rate of the instability
\begin{equation}
\sigma(q)=\sigma_\mathrm{th}\frac{q(2q_\mathrm{th}-q)}{q_\mathrm{th}^2}\label{eq:heuristic}
\end{equation}
leads to a growth of the front corrugation amplitude~$A$ depending on the wavenumber~$q$. Here, the function~$\sigma(q)$ has been chosen as simple as possible, with the condition that it reaches its maximum value~$\sigma_\mathrm{th}$ for $q=q_\mathrm{th}$, where~$q_\mathrm{th}$ is the expected wavenumber from the theory developed by Hassanzadeh \textit{et al.}~\cite{Hassanzadehetal2007} and $\sigma_\mathrm{th}$ is the theoretical growth rate (see Eq.~(\ref{eq:GrowthRateTh})) obtained by Elenius \& Johannsen~\cite{EleniusJohannsen2012}. Note that Eq.~(\ref{eq:heuristic}) also imposes that~$\sigma(q=0)=0$, as in the classical Rayleigh-Taylor instability~\cite{Chandrasekhar1961}. The dependency of the heuristic growth rate~(\ref{eq:heuristic}) with respect to the horizontal wavenumber~$q$ is shown in Figs.~\ref{fig:spectrum}(a) and~(d), for experiments respectively at high and low CO$_2$ partial pressures, already discussed in Figs.~\ref{fig:front_characteristics} and~\ref{fig:lambda}. The growth rate is null for $q=0$, then becomes positive for small~$q$ before reaching its maximum at $q=q_{th}$ and finally decreasing to become negative at large~$q$. Note that the wavenumber range where this function is shown correspond to the typical range in the experiments, i.e. from $q_L=0.021$~mm$^{-1}$ corresponding to the length of the tank~$L=300$~mm to $q_d=4.6$~mm$^{-1}$ corresponding to the typical grain size~$d=1.5$~mm. The spectrum of the front corrugation amplitude~$A$, shown in Figs.~\ref{fig:spectrum}(b) and~(e), is very broad during the initial forcing by the porosity fluctuations because this forcing has been assumed independent of the wavenumber~$q$. However, at later times, the instability still selects the most unstable wavenumber~$q_{th}$ predicted by the theory, i.e. the one corresponding to the maximum of the growth rate. The time dependency of the front corrugation amplitude~$A$ is shown for different wavenumbers in Figs.~\ref{fig:spectrum}(c) and~(f). It is visible that the amplitude of the mode with the wavenumber~$q_{th}$ expected from the theory is slightly larger than the others, but remains on the same order of magnitude for a large range of wavenumbers, at least between the blue and red vertical dashed dotted lines. This may explain the multi-scale nature of the front observed in the experiments within the different scales expected at this CO$_2$ partial pressure range. For higher wavenumbers with negative growth rate, the initial linear growth is clearly visible but rapidly saturates. This saturated value is not completely negligible, as the front corrugation amplitude~$A$ tends to $-D_0/(d \sigma)>0$ when $t\rightarrow \infty$ and $\sigma<0$ (see Eq.~(\ref{eq:amp_linear})), but remains much below the curves obtained for the wavenumbers with positive growth rate. The comparison with the experimental data for these two experiments plotted in Figs.~\ref{fig:spectrum}(c) and~(f) shows relatively good agreement. This heuristic model could therefore explain the global selection of the wavelength by the instability, even in the presence of a broad forcing due to porosity fluctuations.

\section{CO$_2$ flux} \label{sec:Flux}

\subsection{Experimental results}

\begin{figure}[t!]
	\begin{centering}
		\includegraphics[width=1\textwidth]{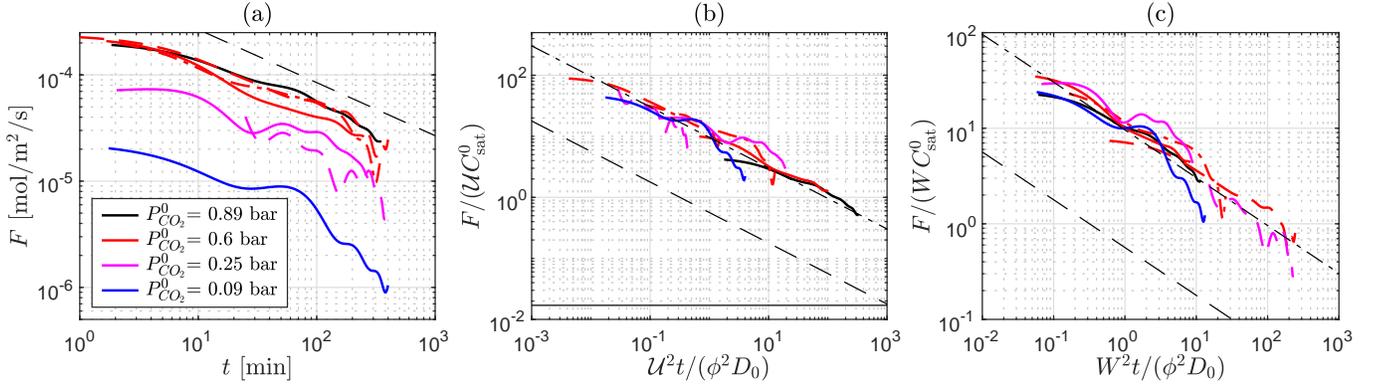}
		\caption{CO$_2$ flux~$F$ as a function of time~$t$ for different initial CO$_2$ partial pressures~$P^0_{CO_2}$ and different permeabilities. The experiments with FEP grains are represented with solid lines, while the ones with silica and PMMA grains are plotted with dashed and dashed dotted lines respectively. The dimensional flux is shown in panel (a), while it has been made dimensionless in panels (b) and (c), using  respectively~$\mathcal{F}=\mathcal{U}C^0_\mathrm{sat}$ and~$\mathcal{F'}=W C^0_\mathrm{sat}$. The time~$t$ has also been made dimensionless, using~$\mathcal{T}=\phi^2D_0/\mathcal{U}^2$ and~$\mathcal{T'}=\phi^2D_0/W^2$. The colors detailed in the legend in panel (a) are the same for all panels. In panel~(b), the horizontal solid line indicates $F/\mathcal{F}=1.7 \times 10^{-2}$, the expected dimensionless flux value for \textit{constant flux regime} once the convection is well established~\cite{Slim2014,DePaolietal2017,Slimetal2013,Pauetal2010}. In panels (b) and (c), the dashed lines represent the expected dimensionless diffusive flux (see Eq.~(\ref{eq:dim_diff_flux})) while the dashed dotted lines highlight a dimensionless diffusive flux with an effective diffusion coefficient $D_\text{eff}=290D_0$.\label{fig:flux_norm}}
	\end{centering}
\end{figure}

This section deals with the CO$_2$ dissolution flux~$F$ in the experiments, measured using the temporal decay of the CO$_2$ partial pressure in the gas compartment above the porous medium (see Section~\ref{flux_measure}). The behavior of the flux~$F$ as a function of time has already been briefly shown and discussed in Fig.~\ref{fig:front_flux_fingers}(e). We here discuss more quantitatively the measured CO$_2$ flux as a function of time for different initial CO$_2$ partial pressures and different permeabilities, first shown in Fig.~\ref{fig:flux_norm}(a) using dimensional quantities. The solid lines correspond to the flux obtained with the FEP grains while the dashed and dashed dotted lines correspond to the flux obtained with the silica and PMMA grains respectively. The colors indicate the initial CO$_2$ partial pressure for the different experiments. All fluxes follow a global decay with time, compatible with a diffusive scaling~$1/\sqrt{t}$ shown by the dashed line. A small kink may be visible at~$t\approx 25$~min for the lowest pressures where the flux grows by about 10 to 30\% before decaying again. This is a typical characteristic of the \textit{flux-growth regime}~\cite{EleniusJohannsen2012,Slim2014,Slimetal2013,DePaolietal2017,Tilton2018,EmamiMeybodi2017}, where the flux is expected to follow the decaying diffusive flux before increasing due to the convection. In our experiments, this increase is much smaller probably because the initial flux is very large as will be shown later. Note that the fluctuations of the flux are also sensitive to the spline interpolation, since the CO$_2$ partial pressure decay is first interpolated to reduce noise before using the time derivative in Eq.~(\ref{eq:def_flux}) which could also explain these small kinks. 

Figure~\ref{fig:flux_norm}(a) clearly proves that the dimensional CO$_2$ flux is larger when the initial CO$_2$ partial pressure is larger, while the variation of the permeability does not seem to influence the flux as it stays on the same order of magnitude for a given CO$_2$ partial pressure. The dependence of the flux with the initial CO$_2$ partial pressure can be understood by the classical theory, where the dissolution flux is usually dimensionalized by~\cite{Slim2014,DePaolietal2017}
\begin{equation}
\mathcal{F}\equiv\mathcal{U} C^0_\mathrm{sat}=\frac{K \Delta \rho g k_H P^0_{CO_2}}{\mu}.
\end{equation}
This characteristic flux is proportional to $(P^0_{CO_2})^2$, since the density difference~$\Delta \rho$ also depends on this pressure (see Section~\ref{sec:equations}). Figure~\ref{fig:flux_norm}(b) shows the CO$_2$ flux~$F$ normalized by the quantity~$\mathcal{F}$ as a function of the dimensionless time~$t/\mathcal{T}=\mathcal{U}^2 t/(\phi^2 D_0)$. It exhibits a relatively good collapse of the different curves obtained at different initial CO$_2$ partial pressures and different permeabilities. However, following the results discussed previously, we can suggest another typical dissolution flux, as the CO$_2$ saturated concentration~$C^0_\mathrm{sat}$ is moved downwards at a constant forcing velocity~$W$. This term can therefore be written as
\begin{equation}
\mathcal{F'}\equiv W C^0_\mathrm{sat}=W k_H P^0_{CO_2}.
\end{equation}
In contrast to $\mathcal{F}$, this typical CO$_2$ forced flux~$\mathcal{F'}$ is proportional to $P^0_{CO_2}$, as~$W$ appears independent of the initial CO$_2$ partial pressure (see Fig.~\ref{fig:front_characteristics}(d)). Note that $W$ has not been measured for the silica and PMMA grains, as the refractive index matching technique with salt water is not suited to these grains. It has therefore been assumed that $W \approx \phi D_0/d$ for these two types of grains. By plotting the normalized CO$_2$ flux~$F/\mathcal{F'}$ as a function of the dimensionless time~$W^2 t/(\phi^2 D_0)$ in Fig.~\ref{fig:flux_norm}(c), we show that all flux curves also collapse well. It thus seems that both scalings are consistent with the experiments. The fact that both these normalization schemes lead to a satisfying collapse of the curves is in fact `natural' because the collapse is independent of the choice of the characteristic velocity $\mathcal{U}_\mathrm{char}$. Indeed, if the flux scales as $1/\sqrt{t}$, $F/\mathcal{F}_\mathrm{char}$ scales as $\sqrt{\mathcal{T}_\mathrm{char}/t}$ with $\mathcal{F}_\mathrm{char}=\mathcal{U}_\mathrm{char} C^0_\mathrm{sat}$ and $\mathcal{T}_\mathrm{char}=\phi^2 D_0/\mathcal{U}_\mathrm{char}^2$. This leads to
\begin{equation}
\frac{F}{\mathcal{U}_\mathrm{char} C^0_\mathrm{sat}}\propto \frac{\phi}{\mathcal{U}_\mathrm{char}} \sqrt{\frac{D_0}{t}},\label{eq:U_char}
\end{equation}
and the characteristic velocity $\mathcal{U}_\mathrm{char}$ disappears from both sides of Eq.~(\ref{eq:U_char}). However, when using $W$ as a characteristic velocity, the dimensionless time~$W^2 t/(\phi^2 D)$ is independent of the CO$_2$ partial pressure, contrary to the one classically used~($t/\mathcal{T}=\mathcal{U}^2 t/(\phi^2 D)$), because $W$ is independent of the initial CO$_2$ partial pressure while $\mathcal{U}$ is linear with this partial pressure. As a consequence, the slight flux growths observed at $t \approx 25$~min in the experiments with the FEP grains stay synchronized using $t/\mathcal{T'}$ in Fig.~\ref{fig:flux_norm}(c), while they are spread horizontally when using $t/\mathcal{T}$ in Fig.~\ref{fig:flux_norm}(b). This therefore shows that the typical CO$_2$ forced flux~$\mathcal{F'}$ may be more relevant to describe the flux obtained in the experiments, in agreement with the forcing of the instability demonstrated previously. Nevertheless, note that the flux curves for the other grains appear shifted as the grain diameter~$d$ is smaller, and $W$ is therefore assumed to be larger. The forcing velocity value for the other grains has however to be checked carefully before being used with trust.

Using both dimensionless fluxes and time scales defined above, the theoretical diffusive flux (see Eq.~(\ref{eq:DiffusiveFlux})) is given by the same formula
\begin{equation}
\frac{F_{diff}}{\mathcal{F}}=\frac{1}{\sqrt{\pi t/\mathcal{T}}}~~~\textrm{or}~~~\frac{F_{diff}}{\mathcal{F'}}=\frac{1}{\sqrt{\pi t/\mathcal{T'}}}.\label{eq:dim_diff_flux}
\end{equation}
It is represented by the dashed lines in Figs.~\ref{fig:flux_norm}(b) and~(c). However, it appears that the typical fluxes obtained in the experiments are about one order of magnitude higher than the expected diffusive flux. This is in clear contradiction with the quantitative fluxes already reported in the literature without grains, which follow the expected diffusive flux well before increasing due to convection~\cite{EleniusJohannsen2012,Slim2014,Slimetal2013,DePaolietal2017,Tilton2018,EmamiMeybodi2017}.

\subsection{Physical discussion}

This one-order of magnitude discrepancy is difficult to explain quantitatively, but may be attributed to flow heterogeneity induced by the granular structure below or close to the Darcy scale, and which cannot be accounted for by the Darcy law as used in many previous works~\cite{EleniusJohannsen2012,Slim2014,Slimetal2013,DePaolietal2017}, despite recent attempts to account for heterogeneities within the porous medium~\cite{Pauetal2010,DePaolietal2016,Tilton2018}. Indeed, using a recent model based on porosity fluctuations~\cite{Tilton2018}, we have shown in Section~\ref{instability_charac} that the grains may directly force the instability and induce an initial flow velocity~$W= D_0 (\partial \phi / \partial z)$. According to this model, the average flux is of second order~$(\varepsilon^2)$ in the asymptotic expansion started in Section~\ref{sec:forcing}. Indeed, the concentration~$c_1$ at first order is supposed to be sinusoidal and the integral over the width of the tank removes its direct contribution to the average flux. The second order~$\varepsilon^2$ is governed by the following equations~\cite{Tilton2018}:
\begin{equation}
\frac{\partial u_2}{\partial x} + \frac{\partial w_2}{\partial z}=0,~~~~~\bm\nabla^2 w_2= \frac{K \Delta \rho g}{\mu}\frac{\partial^2 c_2}{\partial x^2},~~~~~\langle \phi \rangle \frac{\partial c_2}{\partial t} + w_2 \frac{\partial c_b}{\partial z} - \langle \phi \rangle D_0 \bm\nabla^2 c_2 = F_2,\label{eq:order2}
\end{equation}
where $F_2$ is a forcing term equal to
\begin{equation}
F_2=-\langle \phi \rangle \widetilde \phi \,\frac{\partial c_1}{\partial t}-\bm v_1 \cdot \bm\nabla c_1+\langle \phi \rangle D_0 \bm\nabla \left( \widetilde \phi \bm\nabla c_1 \right).
\end{equation}
The equations~(\ref{eq:order2}) are similar to the equations~(\ref{eq:order1}) for the order~$\varepsilon^1$, but the forcing term~$F_2$ is more complex, as it contains more terms and depends on terms of order~$\varepsilon^1$ of the concentration and velocity. Using equations~(\ref{eq:order1}), it can be rewritten as
\begin{equation}
F_2=\left(\langle \phi \rangle D_0 \frac{\partial \widetilde \phi}{\partial z} -w_1 \right) \left(\frac{\partial c_1}{\partial z}-\widetilde \phi \, \frac{\partial c_b}{\partial z} \right) 
+ \left(\langle \phi \rangle D_0 \frac{\partial \widetilde \phi}{\partial x} -u_1 \right) \frac{\partial c_1}{\partial x}.
\end{equation}
Nevertheless, our simple model at order~$\varepsilon^1$ only gives us an estimation of the vertical velocity~$w_1$ at early stages, but not the complete solution of equations~(\ref{eq:order1}). It is therefore difficult to go further analytically using these calculations and a numerical approach as the one provided by Tilton~\cite{Tilton2018} would be necessary. However, it seems that this model cannot explain the large flux measured at early times since $c$ remains small in the linear regime. 

Another mechanism can be thought of to explain the large flux observed in the experiments. Just beneath the surface, the horizontal velocity of the convection cells may induce hydrodynamic transverse dispersion due to the presence of the grains~\cite{Riazetal2006,HidalgoCarrera2009,EmamiMeybodietal2015,EmamiMeybodi2017,EmamiMeybodi2017b,Wenetal2018,Liangetal2018,Souzyetal2020,DePaoli2021}, leading to enhanced vertical diffusion. However, the typical estimation of the hydrodynamic dispersion effects leads to an extra diffusion coefficient~$D_\text{T} = \alpha_\text{T}\langle U\rangle \sim d \langle U\rangle /10$ in the transverse direction~\cite{Souzyetal2020,Liangetal2018}, where $\langle U\rangle$ is the typical mean interstitial velocity just beneath the free surface. Taking $\langle U \rangle = W/\phi \approx D_0/d$ leads to $D_\text{T} \sim D_0/10$, which represents an increase of the diffusion coefficient~$D_0$ by only 10\%. This increase is clearly not sufficient to explain the order of magnitude difference on the flux, i.e. at least two orders of magnitude difference on the diffusion coefficient. Note that hydrodynamic dispersion has been neglected when performing the asymptotic expansion to obtain the forcing terms in Eqs.~(\ref{eq:order1}) and~(\ref{eq:order2}). Indeed, hydrodynamic dispersion is described by an anisotropic tensor (see Eq.~(\ref{eq:def_disp_tens})), and brings a large number of terms at both orders~$\varepsilon^1$ and ~$\varepsilon^2$, that are not easily interpretable. For the sake of clarity, we have therefore derived the model without these terms. However, despite the fact that the transverse dispersion coefficient~$D_\text{T}$ is expected to remain much smaller than~$D_0$, a similar calculation for the longitudinal dispersion coefficient shows that $D_\text{L} = \alpha_\text{L}W/\phi \sim D_0$, i.e. that this effect is on the same order of magnitude as the diffusion coefficient. Hydrodynamic dispersion is therefore not negligible and understanding its effects on the convective dissolution process, with or without a forcing by porosity fluctuations, remains of paramount importance to correctly model geological sequestration sites~\cite{HidalgoCarrera2009,Riazetal2006,EmamiMeybodietal2015,DePaoli2021}. As examples, based on numerical values of structural parameters from Bickle \textit{et al.}~\cite{Bickleetal2007,Bickleetal2017} and Boait \textit{et al.}~\cite{Boaitetal2012}, one can estimate the typical pore scale P\'eclet number $\mathrm{Pe}= U a/D_0\approx U \beta \sqrt{K} /D_0$ in the famous Sleipner and Salt Creek sites to a value between $2\cdot 10^{-2}$ to $5\cdot 10^{-2}$, since the prefactor $\beta$ relating the typical pore size~$a$ to $\sqrt{K}$ amounts to $20-50$ ($U$ being the Darcy velocity). So the typical hydrodynamic longitudinal dispersion normalized by $D_0$, $D_\text{L}/D_0 \simeq \mathrm{Pe} (\alpha_L / a) / \phi$  (with $\alpha_L /a$ typically of about $2$ or $3$ and $\phi \simeq 0.3$), amounts to $\sim 10\,\mathrm{Pe}  = 0.2$ to $0.5$. In our experiments, this quantity is about~$1$, so slightly larger, since the P\'eclet number is $\mathrm{Pe}=W a/D_0\approx 0.1$ when considering $a \simeq 0.3 d$ and $W \sim \phi D_0/d$.

We hypothesize that the CO$_2$ flux measured about one order of magnitude larger than expected may come from the forcing at different scales. Indeed, one can expect the instability to be triggered within each pore close to the surface, leading to a large number of pore-scale plumes not detectable with our measurement techniques. These pore-scale plumes will then merge at larger scales, until the global instability is triggered and measured. The addition of the pore-scale plumes may increase significantly the CO$_2$ flux, but it is necessary to validate quantitatively this mechanism using pore-scale simulations or extra measurements within the pores close to the surface.

To conclude, it is not clear why the flux is so large at early stages. However, it is in agreement with several works using an opaque cell which have measured a fast decay of the CO$_2$ pressure in a gas compartment above a porous medium~\cite{NazariMoghaddametal2012,Seyyedietal2014,NazariMoghaddametal2015}. The authors  have attributed this to convection and have extracted an effective diffusivity about 1 to 2 orders of magnitude larger than the molecular diffusivity, in good agreement with our observations. However, as the experimental cells used were opaque, they were not able to investigate the characteristics of the instability and to connect them with the flux behavior. Note that in the absence of grains, the numerical and theoretical studies have shown that the flux becomes steady at late times (i.e. in the nonlinear regime) and equal to $1.7 \times 10^{-2}\mathcal{F}$~\cite{Slim2014,DePaolietal2017,Slimetal2013,Pauetal2010}. This prediction is plotted as a solid line in Fig.~\ref{fig:flux_norm}(b). It is possible that this regime may be reached at later times and/or at higher CO$_2$ partial pressures in the presence of grains. The steady flux should be dependent on the permeability~$K$, as $\mathcal{F}\propto K$. This is in agreement with the experimental results of Kneafsey and Pruess~\cite{KneafseyPruess2011} in a PVT cell. However, these effects have not been observed here as we focus on the onset of the convective dissolution and at relatively low CO$_2$ partial pressure compared to this reference.

\section{Conclusions}

In this work, we have demonstrated that the combination of refractive index matching and planar laser induced fluorescence can successfully allow  characterizing the onset of the convective dissolution process in a granular, 3-D, porous medium. These quantitative measurements have been completed by flux measurements, using CO$_2$ partial pressure decay in the gas phase above the porous medium. The experimental results obtained have been compared to the predictions mainly obtained using numerical or theoretical approaches assuming Darcy's law in a homogeneous and isotropic porous medium.

The dimensional growth rate~$\sigma$ of the instability has been shown to be constant when the initial CO$_2$ partial pressure increases by a factor 20. This is in clear discrepancy with the theoretical predictions but can be explained by a model based on porosity fluctuations. Indeed, the experimental results for the growth of the front corrugation amplitude are fully compatible with a forcing of the convection by the porosity fluctuations, with a  forcing velocity $W=D_0\partial \phi/\partial z$ found to be on the order of $\phi D_0/d$, i.e. the diffusion coefficient multiplied by the porosity and divided by the typical grain size~$d$ in the experiments. As a consequence, the convection is triggered much faster than expected at low CO$_2$ partial pressures. However, the typical wavelength measured in the experiments remains compatible with the orders of magnitude proposed by the theoretical and numerical approaches without porosity fluctuations, despite a clear multi-scale pattern of the front. This has ben explained by a simple heuristic model showing that the forcing acts at a large range of wavenumbers, and more preferably at large ones, but the wavenumbers growing the fastest are still the ones corresponding to the maximum of the expected growth rate of the instability.

The CO$_2$ flux across the interface has been measured to be at least one order of magnitude higher than expected, compared to the diffusive flux. Despite our efforts, this is not completely explained yet, but the presence of the grains leading to a local forcing must play a large role in this phenomenon. To completely understand this effect, it is necessary to go further, experimentally by directly measuring the flow within the pores close to the surface~\cite{Souzyetal2020}, and/or numerically by using pore scale simulations~\cite{Ramstadetal2019} to investigate such dynamics at small scales. These simulations remain much more costly than the ones performed up to now in the investigation of the convective dissolution instability, but they appear necessary to highlight the local effect of the structure of the pore space at scales smaller than the Darcy scale, instead of assuming Darcy's law in a homogeneous and isotropic porous medium.

The presence of a granular porous medium, intrinsically random at the pore scale, in our experiments, has revealed a completely different onset of the CO$_2$ convective dissolution than the one expected by current numerical and theoretical approaches. In order to understand and predict faithfully the onset of this process in geological sequestration sites, these results show that it is necessary to go beyond the models of flow based on  Darcy's law and to account in some way for the sub-Darcy scale complexity and heterogeneity of the porous media.

\begin{acknowledgments}
This work was carried out in the framework of the {\em CO2-3D} Project (ANR-16-CE06-0001) funded by the French National Research Agency (ANR). The authors thank F. Nadal for fruitful discussions and for providing part of the material used in the experimental setup.
\end{acknowledgments}

\appendix

\section{Measurements of the permeability~$K$ of the different porous media\label{permeability}}

\begin{figure}[t!]
	\begin{centering}
		\includegraphics[width=1\textwidth]{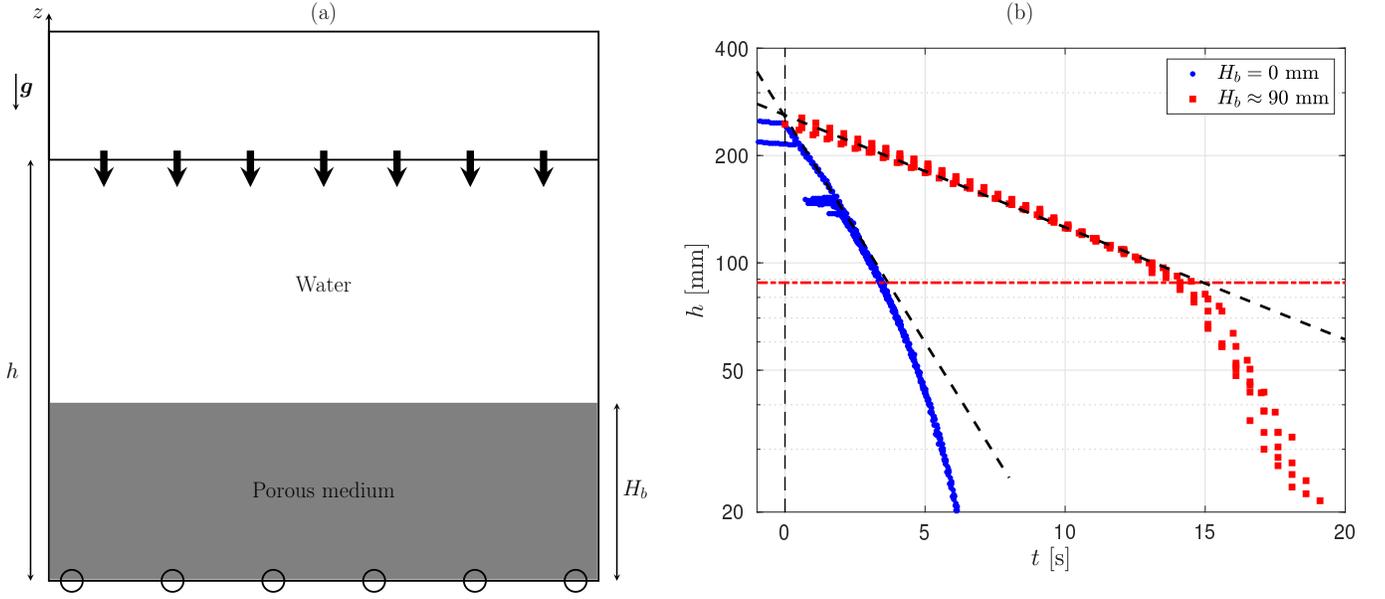}
		\caption{(a)~Sketch of the experimental setup to measure the permeability~$K$ of the porous medium. (b) Data obtained for the FEP grains: water level~$h$ as a function of time in a log-lin scale, without (blue dots) and with (red squares) grains at the bottom of the tank. The two dashed lines show the exponential fits to determine~$\gamma$ (fit of the blue dots) and~$K$ (fit of the red squares).\label{fig:permeability}}
	\end{centering}
\end{figure}

The permeability~$K$ of each porous medium made with the 3 different sets of grains (see Table~\ref{table:extra_grains}) has been obtained by measuring the evolution of the water level in the tank opened at its bottom and containing a porous medium with a given height~$H_b\approx 90$~mm, as illustrated in Fig.~\ref{fig:permeability}(a). It is similar to the one used for the instability characterization, with a typical size of $300 \times 300 \times 15$~mm$^3$. 6 holes are drilled at its bottom: having a diameter of $12$~mm, they are equipped with a small grid to prevent the grains to be evacuated by the flow. Note that the vertical $z$-axis is oriented in the opposite direction of the gravity in that case. With Darcy's law (see Eq.~(\ref{eq:Darcy})) projected vertically, one gets
\begin{equation}
w_P=-\frac{K}{\mu}\left(\frac{\partial P}{\partial z}+\rho_w g\right),
\end{equation}
with $w_P$ being the typical Darcy velocity in the porous medium and $\rho_w$ the water density. At the free surface of the liquid the derivative of $h$ is equal to the fluid velocity $w$ in the tank above the porous medium. The flow-rate conservation at the interface between that tank and the porous medium imposes that 
$w_P = w = \textrm{d}h/\textrm{d}t$, and therefore
\begin{equation}
\frac{\textrm{d}h}{\textrm{d}t}=-\frac{K}{\mu}\left(\frac{\partial P}{\partial z}+\rho_w g\right).\label{eq:dh}
\end{equation}
One can then multiply Eq.~(\ref{eq:dh}) by~$\textrm{d}z$ and integrate it between~$0$ and~$H_b$, i.e. across the porous medium's height. This leads to
\begin{equation}
\frac{\textrm{d}h}{\textrm{d}t}H_b=-\frac{K}{\mu}\left(\int_0^{H_b}\frac{\partial P}{\partial z}\textrm{d}z+\rho_w g H_b\right),\label{eq:dh_int}
\end{equation}
where
\begin{eqnarray}
\int_0^{H_b}\frac{\partial P}{\partial z}\textrm{d}z&=&P(H_b)-P(0)\\
&=&P_\text{atm}+\rho_w g(h(t)-H_b)-\left(P_{atm}-\gamma \frac{\textrm{d}h}{\textrm{d}t}\right)\\
&=&\rho_w g(h(t)-H_b)+\gamma \frac{\textrm{d}h}{\textrm{d}t}.\label{eq:dP_int}
\end{eqnarray}
The last term, $\gamma\, \textrm{d}h/\textrm{d}t$, is here to account for the pressure drop at the bottom of the tank, which is not directly at the atmospheric pressure~$P_\text{atm}$ due to hole configuration and the grids retaining the grains. Combining Eqs.~(\ref{eq:dh_int}) and~(\ref{eq:dP_int}) leads to a differential equation for the water level~$h$
\begin{equation}
\frac{\textrm{d}h}{\textrm{d}t}\left(1+\frac{K \gamma}{\mu H_b}\right)=-\frac{K \rho_w g}{\mu H_b}h(t),\label{eq:dh_final}
\end{equation}
whose solution is given by
\begin{equation}
h(t)=h_i \exp\left(-\frac{K \rho_w g\,t}{\mu H_b+K \gamma}\right),\label{eq:dh_sol}
\end{equation}
with~$h_i$ the initial water level in the tank.

To estimate the pressure drop coefficient~$\gamma$, one first performs experiments without porous medium, such that~$\phi=1$, $H_b=0$ and~$K=e^2/12$. Therefore, the water level~$h$ is expected to behave simply as
\begin{equation}
h(t)=h_i \exp\left(-\frac{\rho_w g\,t}{\gamma}\right).\label{eq:dh_sol1}
\end{equation}
The data obtained for the FEP grains are shown in Fig.~\ref{fig:permeability}(b) with blue dots. They are in good agreement with an exponential decay at early times, i.e. up to~$h=100$~mm where the influence of the position of the holes start to be important. A total of $8$~experiments with different initial water levels~$h_i$ have been performed and a good collapse is observed between these experiments once the initial starting times are adjusted as they start from different water level. The pressure drop coefficient~$\gamma \approx3.37\times 10^4$~kg.m$^{-2}$.s$^{-1}$ is measured by fitting the exponential decay before the water level reaches~$100$~mm. 

Then, the granular porous medium is introduced at the bottom of the tank, with a given height~$H_b\approx 90$~mm. The position of the top of the porous medium is shown in Fig.~\ref{fig:permeability}(b) by a horizontal dashed dotted red line. Several experiments with different preparations of the porous medium (with and without compaction, before or after removing the bubbles) have been performed and do not show any significative difference in the data, all of which are presented in Fig.~\ref{fig:permeability}(b) as red squares. Note that these data have been obtained for the FEP grains, but a similar small dispersion of the data is observable for the two other sets of grains. By fitting the exponential decay above $100$~mm, one gets the permeability of the porous medium consisting of FEP grains as~$K=(9.3\pm 0.8)\times10^{-10}$~m$^2$. The permeabilities for the two other sets of grains are given in Table~\ref{table:extra_grains}, in the main body of the paper.

Note that these values are consistent with the estimation of the permeability from the typical grain size through the Kozeny-Carman equation
\begin{equation}
K=\frac{\phi^3d_p^2}{150(1-\phi)^2},\label{eq:KC}
\end{equation}
with~$d_p$ the diameter of spherical particle of identical volume. For example, for the FEP grains, $d_p\approx2.4$~mm and~$\phi\approx0.39$ (see Section~\ref{sec:PorousMedium}), leading to~$K\approx 6\times10^{-9}$~m$^2$. One therefore recovers the order of magnitude obtained in our experiments, while the limited discrepancy between the two values may come from the fact that the grains we used are not spherical. In addition, we also remind that the prefactor~$1/150$ of the Kozeny-Carman equation is not necessarily even expected to provide a perfect match to the permeability of a pack of monodisperse spheres. Indeed, the prefactor that matches the best the permeability values measured for the PMMA and silica grains (than are spherical) is~$2$ to~$4$ times smaller than the one given in Eq.~(\ref{eq:KC}).

\section{Calibration of the fluorescence as a function of the pH\label{pH}}

\begin{figure}[t!]
	\begin{centering}
		\includegraphics[width=0.75\textwidth]{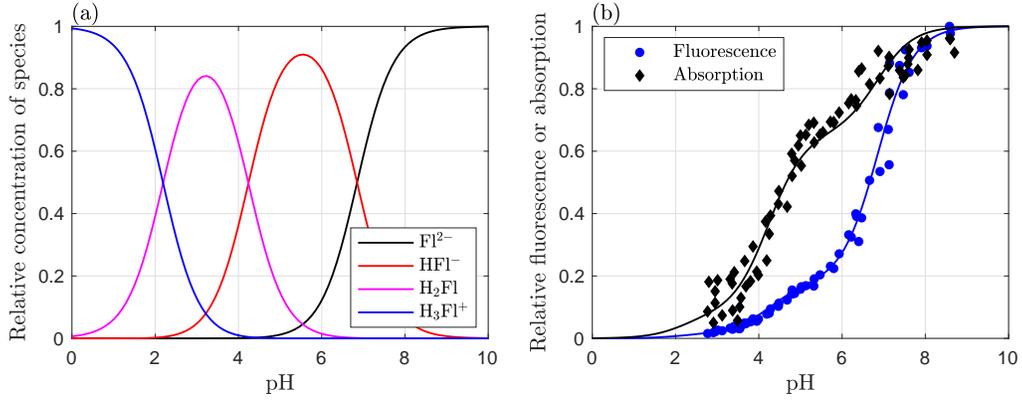}
		\caption{(a)~Bjerrum plot of fluorescein species as a function of pH. (b)~Variation of relative fluorescence (blued disks) and absorption (black diamonds) of a salt water solution of fluorescein as a function of pH. The lines with corresponding colors indicate theoretical curves based on parameters taken from the literature and given in Table~\ref{tab:fluo}.\label{fig:fluo_VS_pH}}
	\end{centering}
\end{figure} 

The total fluorescence~$f(pH)$ of a fluorescein solution at excitation wavelength $\lambda_e$ is given by the sum of the fluorescence~$f_i$ of the individual species~$i$~\cite{DiehlMarkuszewski1989}
\begin{equation}
f(pH)=\sum_{i=1}^4 f_i(pH) \propto \sum_{i=1}^4 \epsilon_i \Phi_i[\textrm{Fl}^i](pH),
\end{equation}
where $\epsilon_i$ is the molar absorption at wavelength~$\lambda_e$, $\Phi_i$ the fluorescence efficiency or quantum yield, and $[\textrm{Fl}^i]$ the concentrations of the different species present in the solution. There is a total of $4$~different species, all connected by 3 acid-base reactions: [Fl$^{2-}$], [HFl$^-$], [H$_2$Fl] and [H$_3$Fl$^+$]. The relative concentration of the different species as a function of pH is shown in Fig.~\ref{fig:fluo_VS_pH}(a), with the dissociation constants indicated in Table~\ref{tab:fluo}. As the anions mostly contribute to the fluorescence~$f$, this quantity therefore varies significantly in the pH range~$4-8$~\cite{MartinLindqvist1975,Walker1987,DiehlMarkuszewski1989,KlonisSawyer1996,CoppetaRogers1998}. Using our setup, we have measured the pH-dependency of the salt water solution of fluorescein, by decreasing the pH with addition of concentrated HCl solution. The measured variations of the relative absorption and fluorescence as a function of pH are shown in Fig.~\ref{fig:fluo_VS_pH}(b). Note that these quantities are normalized by their respective values at high pH. There are in a good agreement with the previous works on fluorescein fluorescence~\cite{MartinLindqvist1975,DiehlMarkuszewski1989,KlonisSawyer1996}, as the theoretical lines represent the predictions based on parameters from this literature (see Table~\ref{tab:fluo}). Note that the pKa for the acid-base reaction between Fl$^{2+}$ and HFl$^+$ is slightly higher than the one reported earlier~\cite{DiehlMarkuszewski1989} and may have changed due to the presence of salt in the solution. However, despite a proper calibration of the fluorescence~$f$, it was not possible to measure quantitatively the CO$_2$ concentrations using this technique, as explained in section~\ref{PLIF}.

\begin{table}[h!]
\caption{The different chemical parameters associated to fluorescein species which have been used to plot the theoretical curves and compare with experimental data in Fig.~\ref{fig:fluo_VS_pH}(b). The references from which the parameters have been extracted or compared are indicated in each column. Note that the dissociation constants on each line is the equilibrium constant for the acid-base reaction with the species of the same line and of the line below. 
\label{tab:fluo}
}
\begin{ruledtabular}
\begin{tabular}{lllll}
\textrm{Fluorescein} & Dissociation constant~\cite{DiehlMarkuszewski1989} &
\textrm{Absorption at $\lambda_e=475$~nm~\cite{KlonisSawyer1996}} &
\multicolumn{1}{c}{\textrm{Quantum yields~\cite{KlonisSawyer1996,MartinLindqvist1975}}}&
\textrm{Fluorescence constant~\cite{DiehlMarkuszewski1989}}\\
\textrm{species} & pKa & 
\textrm{$\epsilon_i$~[cm$^{-1}$.(mol.l)$^{-1}$]} &
$\Phi_i$ &
\textrm{$\epsilon_i \Phi_i/$max$(\epsilon \Phi)\times 100$}\\
\colrule
Fl$^{2-}$ & 6.85 & $48649$ & $0.93$ & $100$\\
HFl$^-$ & 4.24 & $31945$ & $0.25$ & $17.65$\\
H$_2$Fl & 2.19 & $3934$ & $0.20$ & $1.74$\\
H$_3$Fl$^+$ & & $966$ & $0.95$ & $2$\\
\end{tabular}
\end{ruledtabular}
\end{table}

\bibliography{CO2_instability}

\end{document}